\documentclass[showpacs, aps, prb, unsortedaddress,showkeys,longbibliography]{revtex4-1}

\usepackage{amssymb}
\usepackage{latexsym}
\usepackage{amsmath}
\usepackage{graphicx}
\usepackage{braket}
\usepackage{multirow}

\usepackage[dvipsnames]{xcolor}
\usepackage{float}
\usepackage[justification=justified,singlelinecheck=false]{caption}
\usepackage{subcaption}
\usepackage[percent]{overpic}

\usepackage[justification=justified,singlelinecheck=false]{caption}

\begin{document}

\title{Local metamaterials and transition layers}

\date{\today}
\author{%
  Kimya Yadollahpour\textsuperscript{1}, 
  Hossein Khodavirdi\textsuperscript{2},
  and Ankit Srivastava\textsuperscript{1},%
}

\affiliation{%
  \textsuperscript{1}Department of Mechanical, Materials, and Aerospace Engineering, Illinois Institute of Technology, Chicago, IL 60616, USA;\\
  \textsuperscript{2}Senior Acoustic Engineer, Crystal Sonic Inc., Phoenix, AZ 85004, USA%
}

\begin{abstract}
In this paper, we elucidate the concept of local acoustic metamaterials. These are composites which exhibit equi-frequency contours (EFC) which correspond to those expected of homogeneous local acoustic media. We show that EFCs for local acoustic media are conics in 2-dimension and quadrics in 3-dimension. In 2-D, the sure signature of negative properties is if the conic is a hyperbola and in 3-D, the sure signature is the presence of hyperboloids. We note that metamaterial coupling (Willis coupling) has the potential of translating these conics and quadrics in the wave-vector plane but that it does not fundamentally change the shape of these geometries. The local effective properties assigned to a composite in such cases are dispersive (frequency dependent) and they satisfy causality considerations. We finally also show that such properties truly characterize the composite in the sense that they can be used to solve scattering problems involving different samples of the composite. We show that this is made possible through the consideration of transition layers. While the sharp-interface model incurs scattering errors exceeding 20\%  at oblique angles, the Drude-layer model restores agreement to within 2\%  without requiring integral-equation or multi-mode expansions, thereby offering a simple yet highly efficient route to accurate scattering predictions in resonant local acoustic metamaterials.

\end{abstract}

\maketitle

\section{Introduction}

 A mechanical wave only interacts with its environment when it can “see’’ the details: its wavelength $\lambda$ must be comparable to the material scale $a$. If $\lambda \gg a$, the wave skips over structure and the medium can be replaced by a simple average of its constituents—classical static homogenization~\cite{placidi2016review,barchiesi2019pantographic}. As $\lambda$ shrinks toward $a$, however, the wave begins to interrogate the architecture: path length within a cell, multiple scattering, and near-field exchange modify how $\omega$ relates to $\mathbf{k}$ and to direction. In an ideal homogeneous isotropic medium the dispersion is $\omega=c|\mathbf{k}|$, but in an architected medium, e.g., a metamaterial, it deforms in ways that must be captured by \emph{dynamic homogenization}. In practice, effective parameters are defined on infinite periodic composites in 1D, 2D, or 3D~\cite{Nemat-Nasser2011HomogenizationMaterials,Nemat-Nasser2011OverallComposites,Srivastava2011OverallComposites}: static in the long-wavelength limit, and—once the microstructure is probed—dynamic via methods like asymptotic multiscale expansions~\cite{bensoussan2011asymptotic,sanchez1980non,bakhvalov2012homogenisation,parnell2006dynamic,andrianov2008higher,craster2009mechanism,craster2010high,antonakakis2013asymptotics,antonakakis2014homogenisation}, and field-averaging/variational formalisms~\cite{willis2009exact,amirkhizi2017homogenization,nemat2011overall,srivastava2012overall,willis1981variational,willis1983overall,willis1984variational,shuvalov2011effective,willis2011effective,norris2012analytical}. These yield frequency-dependent effective properties and, in general, Willis-type couplings~\cite{milton2007modifications,willis2009exact,willis2011effective}, so the description is temporally dispersive and—in some regimes—\emph{spatially} dispersive (nonlocal), enabling phenomena such as hyperbolic propagation, direction-dependent bandgaps, and negative dynamic mass density~\cite{srivastava2015elastic,norris2012analytical,craster2010high,antonakakis2014homogenisation}. Yet parameters obtained in free space need not reproduce scattering by finite samples~\cite{simovski2009material}. As a complementary route, \emph{retrieval} methods infer parameters by matching a finite slab’s reflection/transmission~\cite{el2000metallic,smith2002determination,amirkhizi2017homogenization,amirkhizi2018overall}; however, such fits often lack portability to different thicknesses, angles, or boundary conditions, motivating the careful distinction—central to this work—between media that remain \emph{local} in the dynamic regime and those that are \emph{nonlocal}.

To accurately model scattering at interfaces involving such non-local media, Fredholm integral equations~\cite{kunin2012elastic} can be employed to provide exact formulations, though they are not computationally efficient or generalizable to complex geometries~\cite{cordaro2023solving, hatamzadeh2011numerical}. An alternative framework is offered by Eringen’s micromorphic theory~\cite{eringen2012microcontinuum, eringen2001microcontinuum}, which extends classical continuum mechanics by introducing deformable particles with internal degrees of freedom, allowing it to describe physical phenomena at sub-continuum scales. Mathematically, the micromorphic approach enables a transformation of the integral constitutive relations of non-local elasticity into a differential form~\cite{eringen1983differential}, making it more tractable for boundary value problems. However, this comes at the cost of increased complexity, as it requires solving a system of multiple partial differential equations with many unknowns~\cite{wang2010micromorphic}. More critically, many key design tools in metamaterials—such as transformation acoustics and elastodynamics~\cite{milton2007modifications, norris2009acoustic, norris2011elastic, smith2002determination}—cannot be directly applied to micromorphic formulations, limiting their practical usefulness in wave-based device engineering.

To determine the effective properties of metamaterials using dynamic homogenization techniques, it is essential to define material parameters such that the homogenized response of a composite matches that of the actual structure. Typically, these properties are first calculated for infinite periodic domains and then applied to predict the behavior of finite systems; however, this can lead to inaccuracies because free-space homogenization neglects evanescent modes~\cite{srivastava2014limit,willis2013some,joseph2015reflection,srivastava2015elastic}, causing discontinuities in the normal field components across interfaces~\cite{simovski2009material} and erroneous scattering coefficients. In electromagnetics, this issue has long been addressed via Drude transition layers~\cite{drude1925theory,simovski2009material,strachan1933reflexion,srivastava2017evanescent}, artificial slabs of thickness on the order of one unit cell inserted between the metamaterial and the surrounding medium to mimic evanescent-field effects and ensure a smooth parameter transition. 

In this paper, we propose an acoustic \textit{local} metamaterial whose boundary‐value problems can be solved with the usual pressure–velocity continuity conditions.  By definition, for a given angular frequency~$\omega$ and in‐plane wave-vector components, a local acoustic medium supports exactly two out-of-plane wave-vector solutions whose energy fluxes point in opposite directions across the interface.  To identify the frequency range over which the metamaterial behaves locally, we first compute its equi-frequency contours (EFCs): in two dimensions these must be conic sections (circles, ellipses or hyperbolae), while in three dimensions they must be quadric surfaces.  Once this local regime is confirmed, we analyze scattering from a microscopic point of view with the \texttt{k-Wave} toolbox in MATLAB, which implements a k-space pseudospectral time-domain (PSTD) scheme for solving the full, heterogeneous acoustic wave equations in the time domain~\cite{treeby2010k}.  The simulated fields provide an accurate basis set that automatically includes propagating and evanescent components, allowing us to enforce boundary continuity without the limitations of analytical plane-wave expansions.  From a macroscopic perspective, dynamic homogenization can underestimate the role of evanescent modes near resonances, leading to discontinuities in the normal field components and, consequently, incorrect reflection and transmission coefficients.  To remedy this, we incorporate \emph{Drude transition layers}: thin artificial slab inserted between the homogenized metamaterial and the surrounding homogeneous medium that emulate the missing evanescent behaviour and restore the correct boundary conditions.  This combined approach—local-regime identification through EFCs, full-wave \texttt{k-Wave} simulations for microscopic scattering, and Drude layers for boundary smoothing—yields more reliable effective parameters and scattering predictions for acoustic metamaterials.

In what follows, we first define local acoustic metamaterials and discuss the characterization of media where interface scattering problems can be solved uniquely using usual acoustic continuity conditions in Section II. In Section III, we focus on the equi-frequency surfaces of local acoustic metamaterials, detailing the equations that define these surfaces and their physical significance in both 2D and 3D contexts. Section IV presents the analysis of local acoustic metamaterial composites, exploring specific examples and unit cell designs, and highlighting the emergence of equi-frequency contours in periodic composites. Finally, we introduce the concept of transition layers and their role in addressing boundary condition issues, particularly through the use of Drude transition layers to simulate evanescent wave effects, thereby improving the accuracy of reflection and transmission coefficients in metamaterials.

\section{Local acoustic metamaterials}

\begin{figure}[htp]
    \centering
    \includegraphics[width=0.45\linewidth]{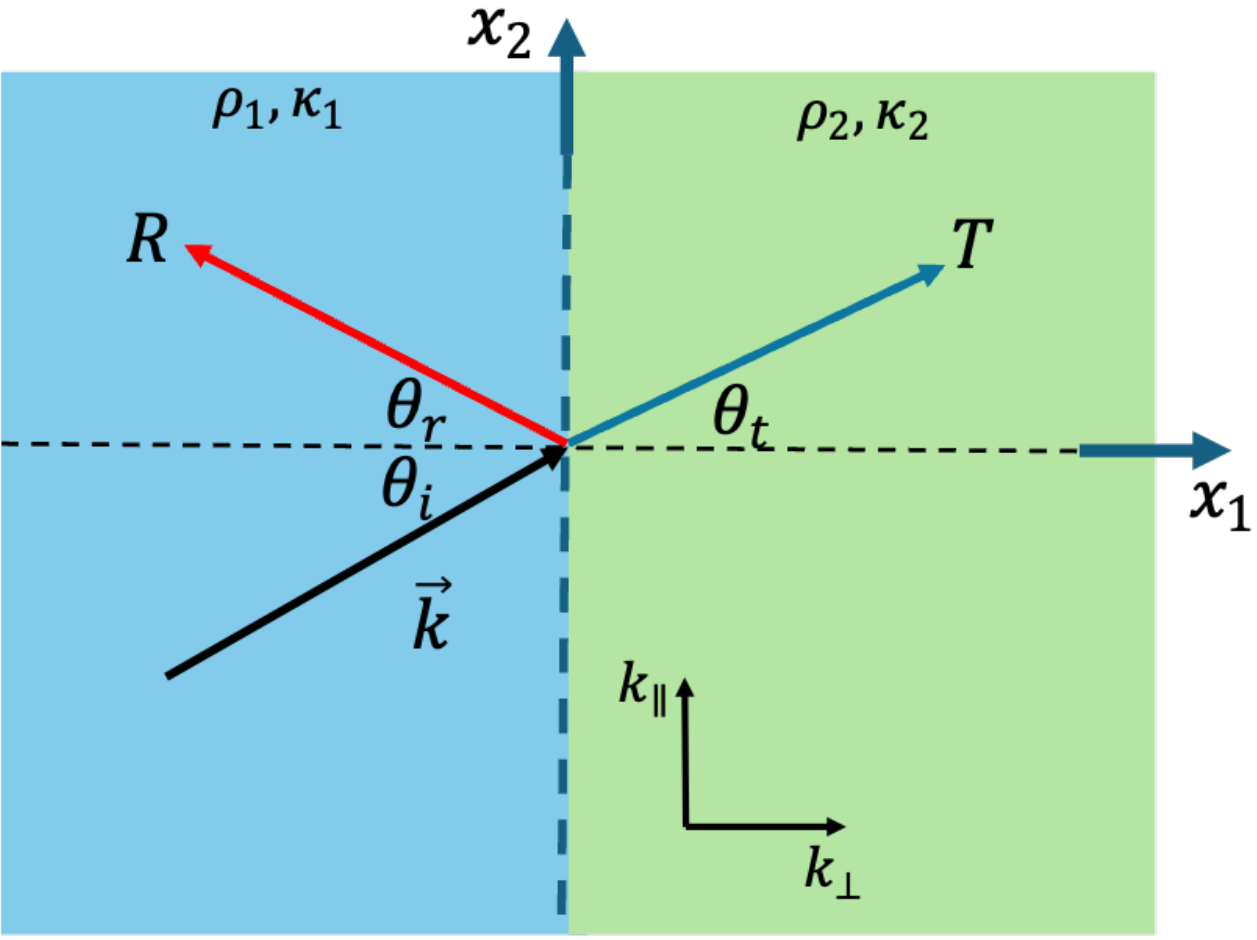}
    \caption{Schematic of two homogeneous acoustic media, characterized by 
\(\rho_1, \kappa_1\) on the left and \(\rho_2, \kappa_2\) on the right, separated by an interface at \(x_1 = 0\).}
    \label{fig:two_homogenous_acoustic_media}
\end{figure}

We will define local metamaterials through the use of interface problems. Our main concern is the characterization of media where interface scattering problems can be solved uniquely using the usual acoustic continuity conditions. We will term such media local - in such media there is no need to invoke additional boundary/interface conditions than what we usually deal with. The field variables of interest in acoustics are pressure $p$ and particle velocity vector $v_i$ where $i=1,2$ in 2D and $i=1,2,3$ in 3D. Consider an interface between two acoustic media (Fig.~\ref{fig:two_homogenous_acoustic_media}) oriented along $x_1=0$. We assume that the two media are characterized by homogeneous material properties such that plane waves of the form $A\exp(i[k_ix_i-\omega t])$ are admissible in each - it should be noted that $A$ is some constant, $\omega$ is the radial frequency, and $\mathbf{k}$ is the wave-vector. We will also occasionally express the wave-vector as $\mathbf{k}=k\mathbf{n}$ where $\mathbf{n}$ would be a unit vector and $k$ would be the wave-number. For a scattering problem at the interface, we have the following information:

\begin{itemize}
    \item Pressure $p$ is continuous
    \item The component of the velocity normal to the interface $v_\perp$ is continuous
    \item Tangential components of the wave-vector interacting with the interface are conserved - this is a restatement of Snell's law
\end{itemize}

For multiple planes waves interacting at the interface, the above three conditions are just enough to determine all the scattering parameters uniquely if only two plane waves (in addition to the incident plane wave) are involved - one in each media. These are the reflected and transmitted waves and the problem of solving the scattering problem involves finding the solutions of the reflected and transmitted wave coefficients. If we decompose the wave-vector into a vector parallel to the interface, $\mathbf{k}_{||}$, and a component perpendicular to the interface, $k_\perp$, then the above requirement is equivalent to demanding that for a given set of $\omega,\mathbf{k}_{||}$, one should have a single $k_\perp$ in either media whose energy flux is either away from the interface or is fully along the interface. This follows from causality and is needed for satisfying the Sommerfeld radiation condition. In general, we expect that for a given $\omega,\mathbf{k}_{||}$, there are two $k_\perp$ solutions with energy fluxes in the $+\perp$ and $-\perp$ directions.

\begin{figure}[htbp]
  \centering
  \begin{overpic}[width=\textwidth,grid=false]{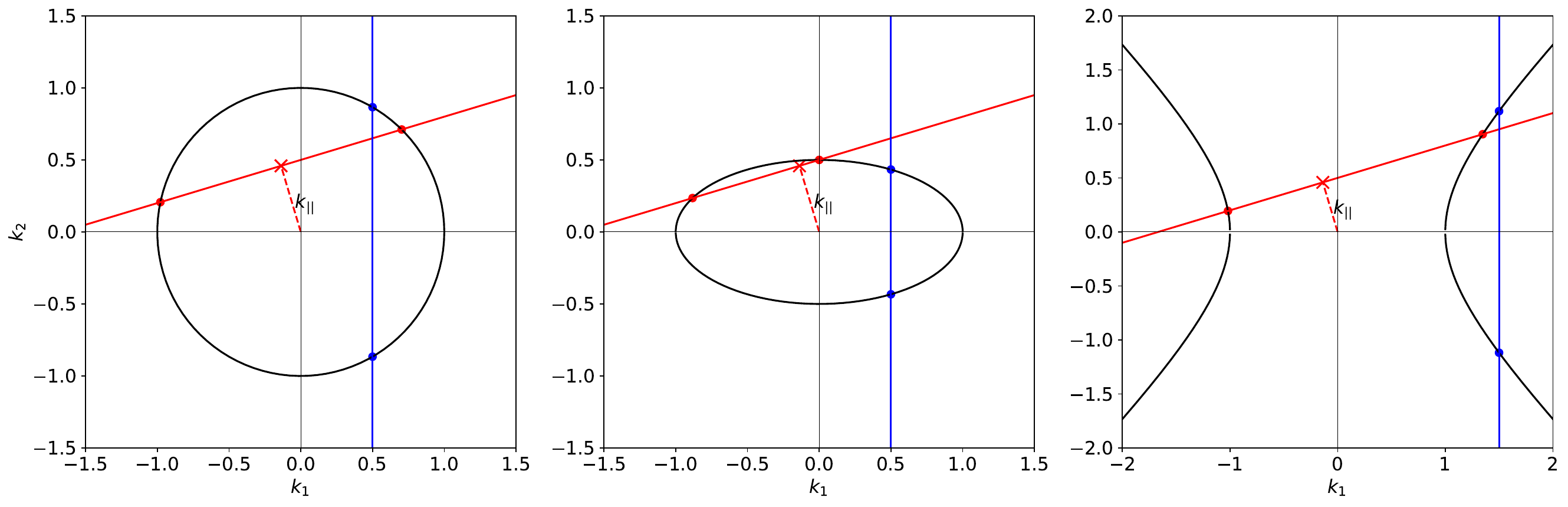}
    \put(-1,32){(a)}
    \put(33,32){(b)}
    \put(66,32){(c)}
  \end{overpic}
  \caption{Illustrations of the three fundamental conic sections, with the red line indicating the tilted interface in each subplot. (a) Circular equi-frequency contour at the origin on the $k_1$--$k_2$ axes; (b) elliptical equi-frequency contour centered at the origin, with principal axes along $k_1$ and $k_2$; and (c) hyperbolic equi-frequency contour centered at the origin, with axes aligned along $k_1$ and $k_2$.}
  \label{fig:conic}
\end{figure}
These ideas can be fixed further by resorting to the concept of equi-frequency surfaces or slowness curves. These are equations of the form $f(\omega,\mathbf{k})=0$ which, when analyzed for constant values of $\omega$, result in all admissible values of $\mathbf{k}$ in the media at the given frequency. If the interface in {Fig.~\ref{fig:two_homogenous_acoustic_media} were to exist in a 2D problem ($x_1-x_2$ plane), then an equi-frequency surface in the form of a conic section:

\begin{align}\label{eq:conic}
Ak_1^2 + Bk_1k_2 + Ck_2^2 + Dk_1 + Ek_2 -\alpha\omega^2=0
\end{align}

is the most general equation which results in a pair of $k_1$ solutions for a given set of $\omega,k_2$. We note that in the above, $\alpha>0$. To make sure that we get a pair of solutions, we note that physical media should exclude the cases where the standard form of the equation above represents a parabola, or curves of order less than 2. This implies that we exclude cases where $B^2=4AC$ (parabola) or $A,C=0$. Fig.~\ref{fig:conic} shows the three physical conic sections (circle, ellipse, and hyperbola) centered around the origin with the $x,y$ axes oriented along the $k_1,k_2$ directions. It also shows that such figures give wave information about interfaces oriented at arbitrary  directions with representative examples showing the emergence of two real $k_\perp$ solutions for a given $k_{||}$. More generally, it can be shown that a straight line can only intersect a conic at either two distinct real points, one real point (in which case the line is either tangent to the conic or the conic under question is a parabola), or two complex points - there is never a case where there is a real intersection and a complex intersection. To see this, we can consider the parametric form of a line $x=x_0+at,y=y_0+bt$ and substitute in the equation of the conic (\ref{eq:conic}) resulting in:

\begin{align*}
A(x_0 + at)^2 + B(x_0 + at)(y_0 + bt) + C(y_0 + bt)^2 + D(x_0 + at) + E(y_0 + bt) + F = 0 
\end{align*}

This is a quadratic equation in $t$ impying that $t$ either has two real roots, one repeated real root, or two complex conjugate roots. Thus conics represent the family of equi-frequency curves for which one gets two $k_\perp$ solutions for any $\mathbf{k}_{||}$ allowing for the unique solution of the acoustic scattering problem. The analogue of conics in 3D are quadrics. These are equations of the form:

\begin{align*}
Ak_1^2+Bk_2^2+Ck_3^2+Dk_1k_2+Ek_2k_3+FK_3k_1+Gk_1+Hk_2+Ik_3+\alpha\omega^2=0
\end{align*}

Fig. (\ref{fig:quadric}) shows some quadric surfaces which emerge by assuming various values of the coefficients.

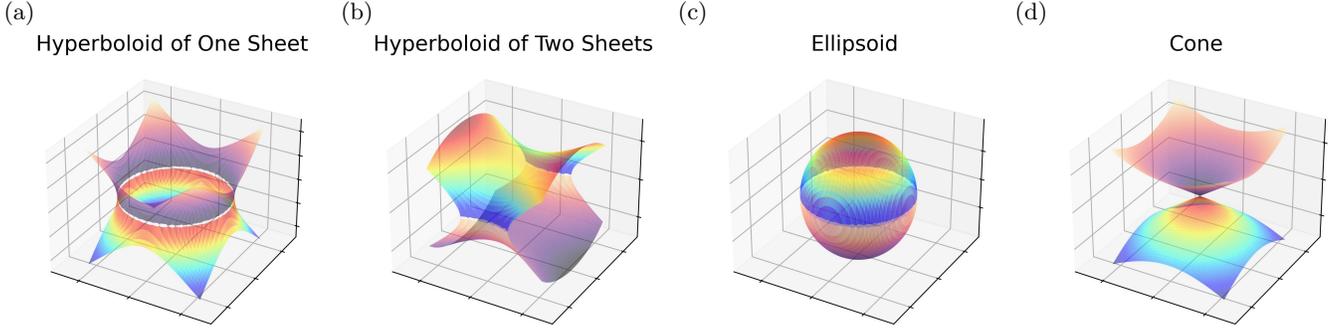
\begin{figure}[htbp]
  \centering
  \begin{overpic}[width=\textwidth,grid=false]{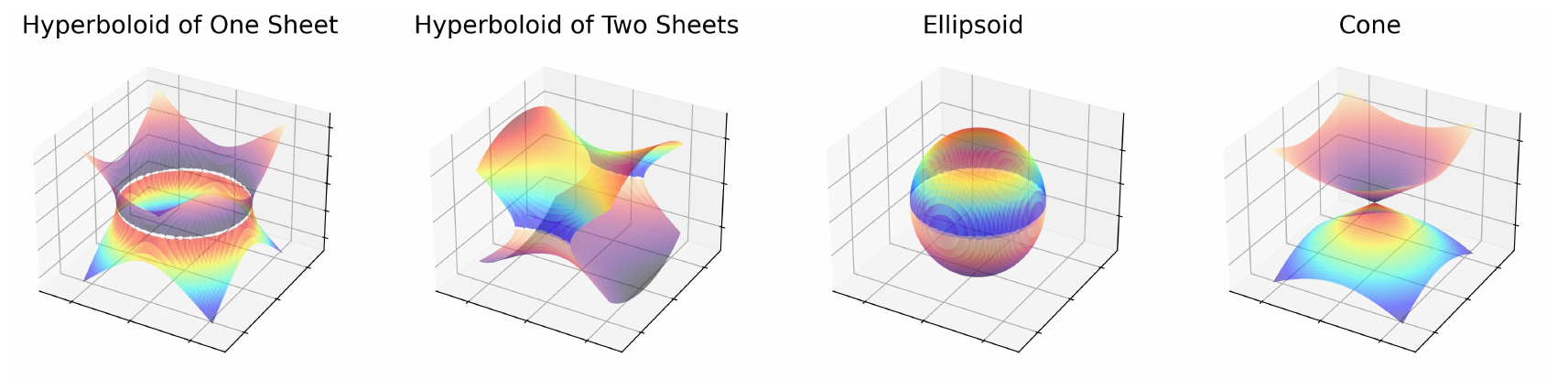}
    \put(-1,25){(a)}
    \put(24,25){(b)}
    \put(49,25){(c)}
    \put(74,25){(d)}
  \end{overpic}
  \caption{3D representations of quadric equi-frequency surfaces:
    (a) One-sheet hyperboloid, arising when one density component is negative;
    (b) Two-sheet hyperboloid, arising when two density components are negative;
    (c) Ellipsoid, when all density components share the same sign;
    (d) Conical equi-frequency surface. }
  \label{fig:quadric}
\end{figure}
\section{Equi-frequency surfaces of local acoustic metamaterials}

Let's consider the acoustic wave problem:

\begin{align*}
\nabla \cdot \left(\frac{1}{\rho}\nabla p\right)=-\omega^2\frac{1}{\kappa}p  \end{align*}

and write the above as:

\begin{align*}
\nabla \cdot  \left(\frac{1}{i\omega}\frac{1}{\rho}\nabla p\right)=i\omega\frac{1}{\kappa}p 
\end{align*}

We can decompose the above into two generalized relations:

\begin{align*}
i\omega v_i = \bar{\rho}_{ij}p_{,j},\quad i\omega p=\kappa v_{i,i}
\end{align*}

Seeking solutions of the form $\psi=\bar{\psi}\exp(ik_ix_i)$ where $\bar{\psi}$ is a constant and $\mathbf{k}$ is the wave-vector, we get:

\begin{align*}
\omega \bar{v}_i = k_j\bar{\rho}_{ij}\bar{p},\quad \omega \bar{p}=k_i\kappa \bar{v}_{i}
\end{align*}

resulting in a single equation:

\begin{align*}
\omega^2\delta_{ij} \bar{v}_j=\bar{\rho}_{ij}\kappa k_jk_o\bar{v}_o
\end{align*}

We are interested in determining the shape of the equi-frequency surfaces for such a material. This involves finding the solutions $\omega(\mathbf{k})$ and can be done by identifying the RHS of the above equation with a square matrix with components $A_{ij}=k^2\kappa\bar{\rho}_{io}n_on_j$. The problem has non trivial solutions when $\mathrm{det}(\omega^2\delta_{ij}-\bar{\rho}_{io}k_ok_j)=0$ which results in the relation:

\begin{align}\label{eq:EFCacoustics2D1}
\kappa\bar{\rho}_{11}k_1^2 + \kappa\bar{\rho}_{22}k_2^2 +\kappa(\bar{\rho}_{12} + \bar{\rho}_{21})k_1k_2-\omega^2=0
\end{align}

The above is the generalized relation of a conic section as discussed earlier. The simplest case is if $\bar{\rho}_{ij}=\rho\delta_{ij}$ in which case the conic section is a circle. Thus, regular isotropic acoustic media exhibits circular equi-frequency contours. Circular contours are also exhibited by double negative isotropic media where $\bar{\rho}_{ij}=\bar\rho\delta_{ij}$ with $\bar\rho<0$ and $\kappa<0$. Still focusing on the diagonal case, if we introduce anisotropy by insisting that $\bar\rho_{11}\neq\bar\rho_{22}$ then we get equi-frequency contours which are elliptic for both regular anisotropic media and double negative anisotropic media. 

\vspace{0.4\baselineskip} 
\vspace{0.4\baselineskip} 
\vspace{0.4\baselineskip} 
\vspace{0.4\baselineskip} 
\vspace{0.4\baselineskip} 

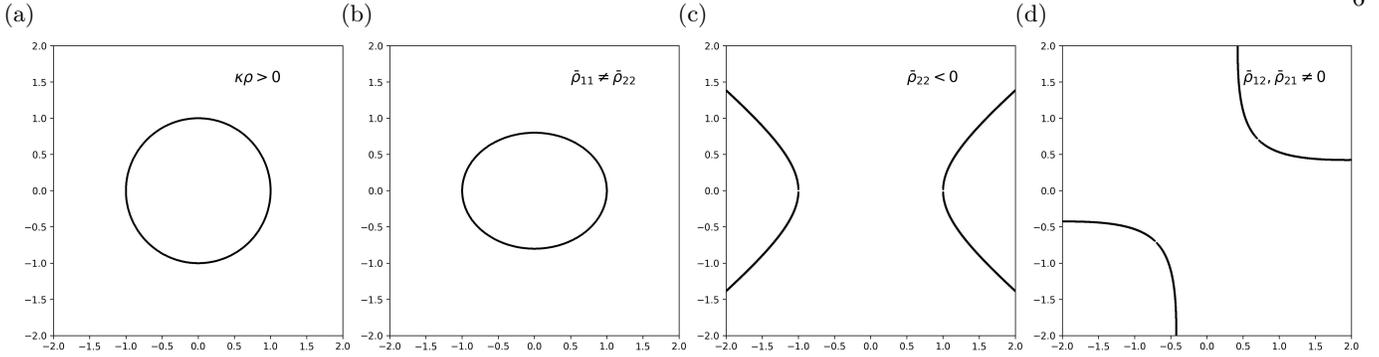
\begin{figure}[htp]
\centering
  \begin{overpic}[width=\textwidth,grid=false]{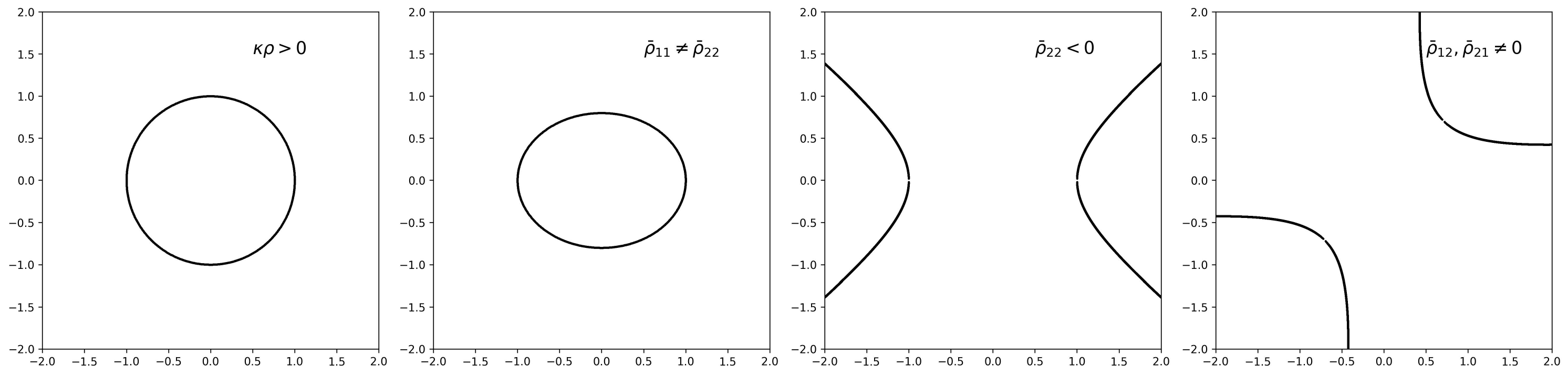}
    \put(-1,25){(a)}
    \put(24,25){(b)}
    \put(49,25){(c)}
    \put(74,25){(d)}
  \end{overpic}
\caption{ Equi-frequency contours of local acoustic metamaterials in 2D. (a) A circular EFC in which the media is isotropic and $\kappa\rho > 0 $. (b) An elliptical equi-frequency contour where the medium is anisotropic, with $\bar\rho_{11} \neq \bar\rho_{22}$. (c) A hyperbola EFC with $\bar\rho_{22}<0$ and $\kappa, \rho>0$. (d) A translated Equi-frequency contour in which there exists $\bar\rho_{12}$ and $\bar\rho_{21}$.}
\label{fig:localAcoustic2D}
\end{figure}
If we now make $\bar\rho_{22}<0$ while keeping $\kappa,\bar\rho_{11}$ positive, then we arrive at hyperbolic equi-frequency contours. In fact, hyperbolic equi-frequency contours are maintained when just one of $\bar\rho_{11},\bar\rho_{22}$ has a different sign that the other two. We note, as a corollary, that circular or elliptical contours may or may not suggest the presence of negative properties, but hyperbolic contours definitely point towards the existence of negative properties. The coefficients of $k_1k_2$ in (\ref{eq:EFCacoustics2D1}) do nothing but rotate the conic about the $k_1,k_2$ axis and, as such do not add any fundamental nuance to the discussion. Their effect is merely to change the primary axes of anisotropy, if it exists. It is possible for $\bar\rho_{12},\bar\rho_{21}$ to be complex as long as they are complex conjugates of each other. In that case, the sum $(\bar{\rho}_{12} + \bar{\rho}_{21})$ is still real and the solutions to the conic equation are still to be found in the real $k_1,k_2$ plane. In, fact $\bar\rho_{12}=\bar\rho_{21}^*$ is demanded by passivity requirements in any case.


In 3D, we have the relation:

\begin{align}\label{eq:EFCacoustics3D1}
\kappa\bar\rho_{11}k_1^2 + \kappa\bar\rho_{22}k_2^2 + \kappa\bar\rho_{33}k_3^2 + \kappa (\bar\rho_{12}+\bar\rho_{21})k_1k_2 + \kappa (\bar\rho_{23}+\bar\rho_{32})k_2k_3 + \kappa (\bar\rho_{12}+\bar\rho_{21})k_1k_2-\omega^2=0
\end{align}

which is the equation of a quadric centered at the origin. Let's first consider the case when the specific density tensor is diagonal. When all the diagonal components are equal, (\ref{eq:EFCacoustics3D1}) is a sphere. This is also true if the diagonal components are equal and negative provided that in that case $\kappa$ is also negative. Thus, equi-frequency surfaces of isotropic regular and isotropic double negative acoustic media are spherical. When the diagonal components are made different (while keeping them positive), then the equi-frequency surfaces are ellipsoidal. This is also true for fully double negative anisotropic and diagonal specific density tensor. The various surfaces which appear in the case of a fully positive or double negative media are shown in Fig. (\ref{fig:doubleNegative}). 

\begin{figure}[htp]
\centering
  \begin{overpic}[width=\textwidth,grid=false]{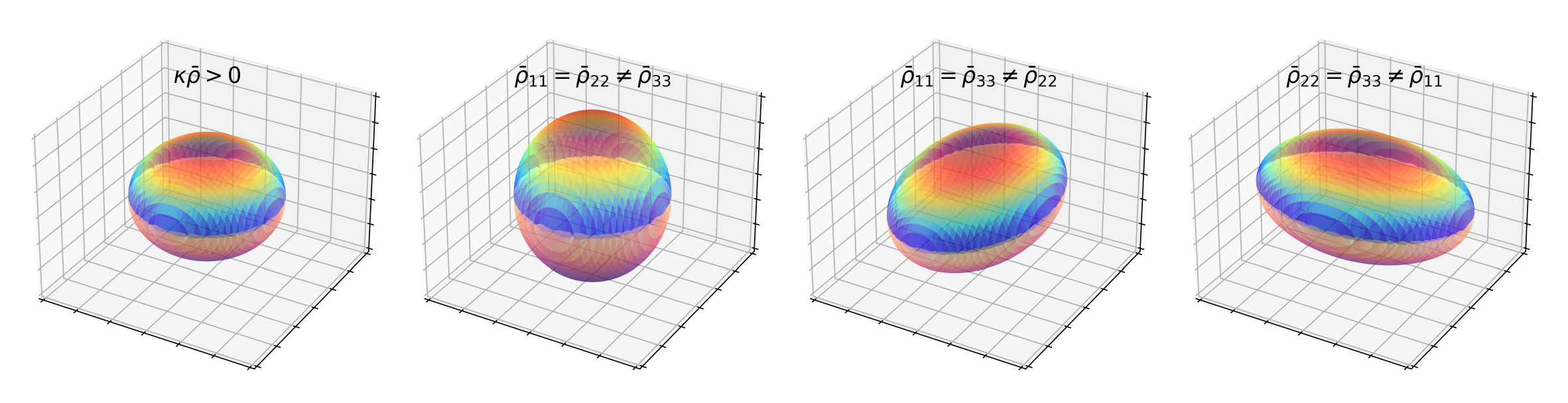}
    \put(-1,25){(a)}
    \put(24,25){(b)}
    \put(49,25){(c)}
    \put(74,25){(d)}
  \end{overpic}
\caption{3D Generalized quadric surfaces in which the density tensor is diagonal. (a) A spherical EFS where all the diagonal components are equal. (b) An ellipsoidal EFS with positive or double negative diagonal components, but $\bar{\rho}_{33}$ has a different value from $\bar{\rho}_{11}$ and $\bar{\rho}_{22}$ ($\bar{\rho}_{11} = \bar{\rho}_{22} \neq \bar{\rho}_{33}$). (c) An ellipsoidal EFS with positive or double negative diagonal components, but $\bar{\rho}_{22}$ has a different value from $\bar{\rho}_{11}$ and $\bar{\rho}_{33}$ ($\bar{\rho}_{11} = \bar{\rho}_{33} \neq \bar{\rho}_{22}$). (d) An ellipsoidal EFS with positive or double negative diagonal components, but $\bar{\rho}_{11}$ has a different value from $\bar{\rho}_{22}$ and $\bar{\rho}_{33}$ ($\bar{\rho}_{22} = \bar{\rho}_{33} \neq \bar{\rho}_{11}$).
}
\label{fig:doubleNegative}
\end{figure}
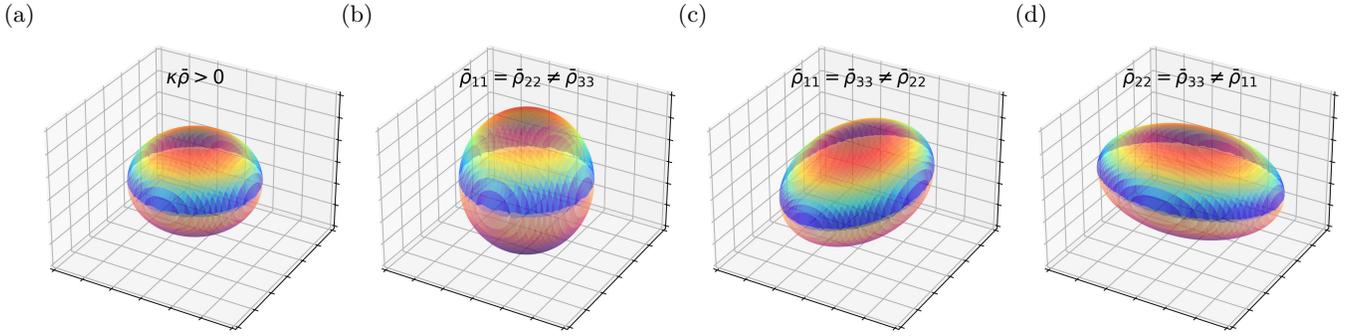

When precisely one of $\bar\rho_{11},\bar\rho_{22},\bar\rho_{33}$ is negative, while the other two remain positive, the resultant quadric surface is a hyperboloid of one sheet. In any orthogonal plane slicing the principal axis, the surface section manifests as an ellipse, whereas sections orthogonal to this plane produce hyperbolas. When any two of $\bar\rho_{11},\bar\rho_{22},\bar\rho_{33}$ are negative, with the remaining coefficient being positive, the quadric surface in question is a hyperboloid of two sheets. As with its single-sheeted counterpart, orthogonal planes intersecting the primary axis render elliptical sections, while those orthogonal to it produce hyperbolas. The various surfaces which appear in the case of an acoustic media with not all specific density components negative are shown in Fig. (\ref{fig:hyperboloid3D}). As with the 2D case, we note that spherical or ellipsoidal surfaces may or may not suggest the presence of negative properties, but hyperboloid surfaces definitely point towards the existence of negative properties. The effect of the off-diagonal terms in the specific density tensor is merely to rotate the specific quadric determined by $\bar\rho_{11},\bar\rho_{22},\bar\rho_{33}$ in the 3-dimensional space. They do not add any new fundamental property to the equi-frequency contours.

\begin{figure}[htp]
\centering
  \begin{overpic}[width=\textwidth,grid=false]{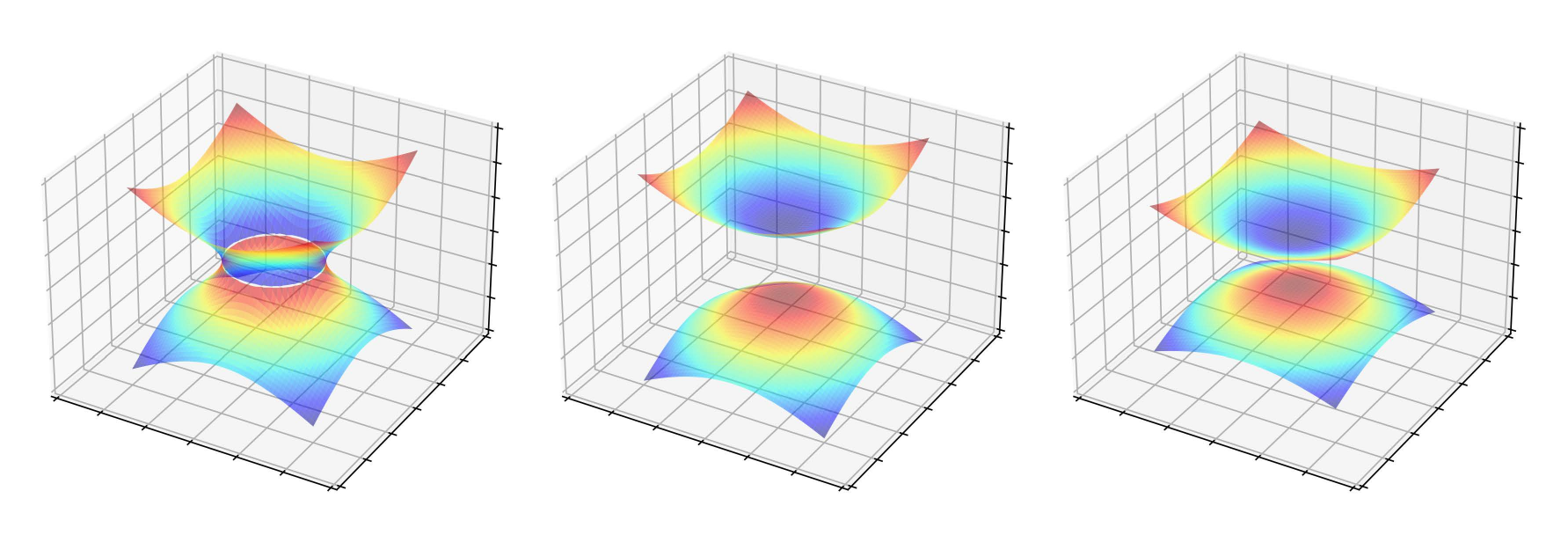}
    \put(-1,32){(a)}
    \put(33,32){(b)}
    \put(66,32){(c)}
  \end{overpic}
\caption{(a)3D EFS representation of one sheet hyperboloid . (b) A hyperboloid of two sheets where $\bar\rho_{11}, \bar\rho_{33}<0$ and $\bar\rho_{22} >0$. (c) A hyperboloid of two sheets where $\bar\rho_{22}, \bar\rho_{33}<0$ and $\bar\rho_{11} >0$.}
\label{fig:hyperboloid3D}
\end{figure}

At this point, one may ask: what is the effect of additional constitutive parameters on the equi-frequency contours? Let's, for example, consider the more general coupled relation which appears in literature under the name of Willis tensor \cite{willis2009exact,milton2007modifications,willis2011effective:}


\begin{align}\label{eq:acousticWillis}
i\omega v_i = ia_i p + \bar{\rho}_{ij}p_{,j},\quad v_{i,i}=i\omega p/\kappa + b_ip_{,i}
\end{align}

where $a_i,b_i$ are new constitutive parameters. The above, under plane wave assumption becomes:

\begin{align*}
\omega \bar v_i = a_i \bar p + k_j\bar{\rho}_{ij}\bar p,\quad k_i\bar v_{i}=\omega \bar p/\kappa + k_ib_i\bar p
\end{align*}


The above can be used to eliminate $\bar p$ and write a single homogeneous equation in terms of $\bar v$:

\begin{align*}
\omega\delta_{ij} \bar v_j = \frac{\kappa}{\omega + \kappa k_mb_m}(\delta_{io}a_ok_j + \bar{\rho}_{io}k_ok_j)\bar v_{j}
\end{align*}

For nontrivial solutions, we set the determinant of the associated matrix equal to 0:

\begin{align}\label{eq:det}
\mathrm{det}\left(\omega\delta_{ij} - \frac{\kappa}{\omega + \kappa k_mb_m}(\delta_{io}a_ok_j + \bar{\rho}_{io}k_ok_j)\right)=0
\end{align}

This results in the following expression:

\begin{align}\label{eq:EFCacoustics2DWillis}
\kappa\bar{\rho}_{11}k_1^2 + \kappa\bar{\rho}_{22}k_2^2 +\kappa(\bar{\rho}_{12} + \bar{\rho}_{21})k_1k_2+\kappa(\omega b_1-a_1)k_1+\kappa(\omega b_2-a_2)k_2-\omega^2=0
\end{align}

given that $\kappa(b_1k_1+b_2k_2)+\omega\neq 0$. This is again the general equation of a conic but with the additional linear terms appearing now which did not appear in the case of the uncoupled equations (\ref{eq:EFCacoustics2D1}). The shape and nature of the conic is still determined by $\bar{\rho}_{11},\bar{\rho}_{22}$, precisely as it was determined in the uncoupled case. $\bar{\rho}_{12},\bar{\rho}_{21}$ serve to rotate the conic about the $k_1,k_2$ axis, just as before. The new linear terms have the effect of translating the conic along the $k_1,k_2$ directions. Specifically, if $\omega b_1-a_1\neq 0,\omega b_2-a_2= 0$ then the conic is translated along the $k_1$ axis. Similarly, if $\omega b_1-a_1= 0,\omega b_2-a_2\neq 0$ then the conic is translated along the $k_2$ axis. With both terms being non-zero, the conic is translated in the plane. If the coupling parameters are such that $\omega b_1-a_1= 0,\omega b_2-a_2= 0$, then the equi-frequency contours are indistinguishable from the case when there are no coupling parameters at all. We also note that it is important that the coupling parameters be such that $\omega b_1-a_1,\omega b_2-a_2$ are real quantities. If either of these is a complex quantity, then real propagating wave solutions cannot in general be found in the medium. Analogously to the case when there was no coupling, this analysis also clarifies that the coupling parameters have no role in determining if a material may be termed positive or negative. The material can be understood to have negative properties only if one or both of $\bar\rho_{11},\bar\rho_{22}$ are negative. Furthermore, it is impossible to extract the values of $a_i,b_i$ separately from an experiment which leverages the dispersion properties since these terms appear only in combination with each other in the equi-frequency contours. Any combination of $a_i,b_i$ such that $\omega b_1-a_1,\omega b_2-a_2$ are constants result in precisely the same equi-frequency contours and are, thus, indistinguishable from each other.

\begin{figure}[htp]
\centering
  \begin{overpic}[width=\textwidth,grid=false]{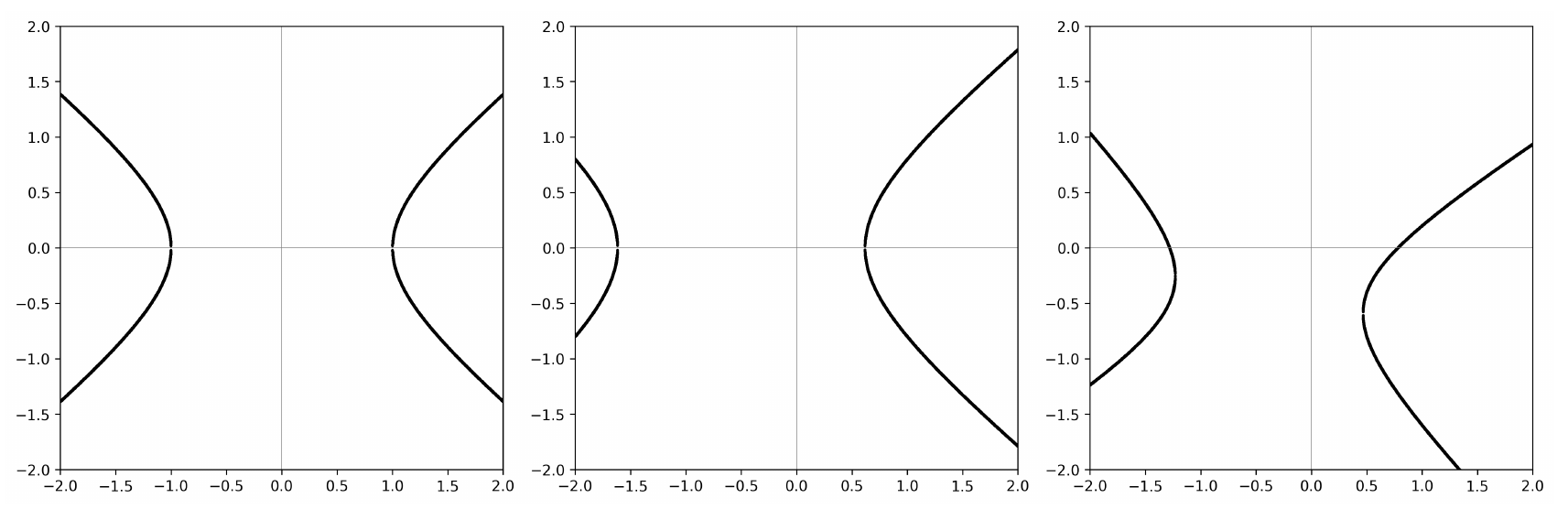}
    \put(-1,32){(a)}
    \put(33,32){(b)}
    \put(66,32){(c)}
  \end{overpic}
\caption{ Impact of coupling parameters on the equi-frequency contours in 2D. (a)A hyperbola EFC with no translation when $\omega b_1- a_1=0$ and $\omega b_2- a_2 =0$. (b) A hyperbola translated along the $k_1$ axis when $\omega b_1- a_1\neq 0$ and $\omega b_2- a_2 =0$. (c) A hyperbola translated along the $k_2$ axis when $\omega b_2- a_2\neq 0$ and $\omega b_1- a_1 =0$.}
\label{fig:WillisAcoustic2D}
\end{figure}

 Figure\ref{fig:WillisAcoustic2D} illustrates how the additional linear terms influence the translation of the conic sections in the equi-frequency contours along the $k_1$ and $k_2$ axes. In Figure\ref{fig:WillisAcoustic2D}b , the translated hyperbola is shown along the $k_1$ axis where $\omega b_1 - a_1 \neq 0$ and $\omega b_2 - a_2 = 0$. The translation of the EFC along the $k_2$ axis, where $\omega b_1 - a_1 = 0$ and $\omega b_2 - a_2 \neq 0$, is depicted in Figure 6c. When we extend this analysis to 3D by solving (\ref{eq:det}), we get a generalized quadric, in a complete analogue of the generalized conic in Eq. (\ref{eq:EFCacoustics2DWillis}). The extra terms which appear, when compared to the uncoupled case (\ref{eq:EFCacoustics3D1}), are linear in $k_i$ and, therefore, serve only to translate the quadric determined by $\bar\rho_{ii}$ in the 3-dimensional space. The amount and direction of translation is determined by $\omega b_i-a_i$, as in the 2D case. There are no fundamentally different observations to be added to the 3D coupled case which were not already mentioned for the 2D coupled case.

\subsection{Partially local acoustic metamaterials}

We recall that we define local acoustic metamaterials as those materials which result in two $k_\perp$ solutions for any $\mathbf{k}_{||}$. In 2D, conics satisfy this property and in 3D, quadrics satisfy it. Real composites generally exhibit dispersion which is more complicated than simple conics or quadrics. Is it possible to assign local properties to such composites then? An example is shown in Fig. \ref{fig:partially_local_metamaterial}a where a hypothetical 2D equi-frequency contour is shown which cannot be approximated by a single conic. The effect of this on wave problems is that there are straight lines which intersect with the curves at more than two points, essentially meaning that for some $\mathbf{k}_{||}$ there are more than two $k_\perp$ solutions. This makes the unique determination of scattering parameters impossible at interfaces which align with these $\mathbf{k}_{||}$ directions. Fig. \ref{fig:partially_local_metamaterial}a shows two such cases where the lines have more than two intersections with the curves. If the interface is aligned along the $\mathbf{k}_{||}$ directions shown and if the incident wave has a tangential wave-number component equal to the $\mathbf{k}_{||}$ shown, then it is impossible to uniquely solve the scattering problem in such cases. It is important to keep in mind that it is not impossible to solve the scattering problem in a composite which exhibits such an equi-frequency contour, but it is impossible to express the composite as a regular acoustic media and solve the problem with the regular boundary conditions available to acoustic media.

\begin{figure}[htp]
  \centering
  \begin{overpic}[width=\textwidth,grid=false]{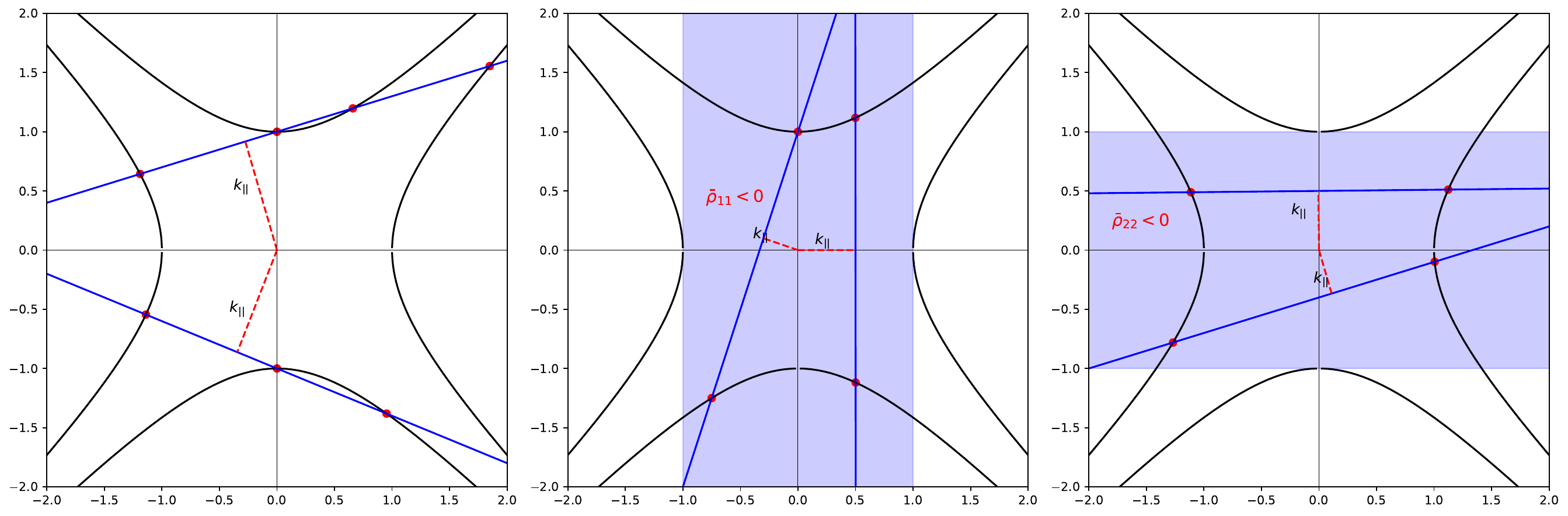}
    \put(-1,33){(a)}
    \put(33,33){(b)}
    \put(66,33){(c)}
  \end{overpic}
\caption{(a) A hypothetical 2D EFC that cannot be approximated by a single conic, indicating non-local behavior. (b) A region where the EFC can be approximated by an anisotropic medium with $\bar{\rho}_{11} < 0$ and $\bar{\rho}_{22} > 0$. (c) A region where the EFC can be approximated by an anisotropic medium with $\bar{\rho}_{11} > 0$ and $\bar{\rho}_{22} < 0$. In both (b), and (c) the shaded area shows the angles of incidence at which the metamaterial behaves locally.}
\label{fig:partially_local_metamaterial}
\end{figure}

We can attempt some resolution to this issue by considering partial regions of the $k_1-k_2$ space which isolate good conic approximations. For example, the equi-frequency contour in the blue region in Fig. \ref{fig:partially_local_metamaterial}b can be approximated well by an anisotropic acoustic media with $\bar\rho_{11}<0,\bar\rho_{22}>0$. As long as the blue lines are oriented such that their intersections all lie in the blue region, the medium behaves precisely as a local acoustic metamaterial $\bar\rho_{11}<0,\bar\rho_{22}>0$ with interface problems solvable with the usual boundary conditions of acoustics. This in turn limits the angles of incidences which are admissible for different orientations of the interface. For example, if the interface is along the $x$ axis (or $k_1$ axis in \ref{fig:partially_local_metamaterial}b), then incidence angles from normal incidence all the way to a maximum incidence angle such that $k_{||}=1.0$ are admissible for the manifestation of local behavior. If we consider interfaces which are at a slight inclination to the $x$ axis then the range of admissible incident angles is curtailed, as can be devised from the inclined line in Fig. \ref{fig:partially_local_metamaterial}b. At the other extreme, when the interface is along the $y$ axis, then the composite again behaves like a local media but now with properties $\bar\rho_{11}>0,\bar\rho_{22}<0$ (Fig. \ref{fig:partially_local_metamaterial}c). Thus, the composite cannot be assigned a single set of local properties (uncoupled or coupled) for all directions of wave travel. But there exist ranges of wave travel directions where the composite can be assigned unique local properties. As long as waves are restricted to these ranges, the composite behaves exactly like a local acoustic media. Finally, there are directions of wave travel where it is impossible to assign any local properties to the composite - examples of these are provided in Fig. \ref{fig:partially_local_metamaterial}a.

\subsection{Effect of nonlocality}

Let's now consider the effect of nonlocality. Let's say that the relevant equations are:

\begin{align*}
i\omega v_i(x) = \int_\Omega\bar{\rho}(x-x')p_{,i}(x')dx',\quad i\omega p(x)=\int_\Omega\kappa(x-x') v_{i,i}(x')dx'
\end{align*}

where $\Omega$ is some volume centered at $x$. We do a Taylor series expansion on $p_{,i},v_{i,i}$ about $x$, explicitly showing only the first two terms:

\begin{align*}
v_i(x)=p_{,i}(x)\int_\Omega\bar{\rho}(x-x')dx' + p_{,ij}(x)\int_\Omega\bar{\rho}(x-x')\bar{x}_jdx'+...\\
p(x)=v_{i,i}(x)\int_\Omega\kappa(x-x')dx' + v_{i,ij}(x)\int_\Omega\kappa(x-x')\bar{x}_jdx'+...
\end{align*}

Relabeling the constitutive parameters and keeping only the first two terms, we have:

\begin{align*}
i\omega v_i=ap_{,i} + b_jp_{,ij},\quad i\omega p=cv_{i,i} + d_jv_{i,ij}
\end{align*}

where $a,b,c,d$ are modified constitutive parameters. When we substitute the traveling wave form in the above and combine to extract the dispersion relation, we will find that the resulting equation is a fourth order polynomial in $k_i$. This results from the second order derivative terms in each of the above equations. In general, therefore, we expect that the kind of weak nonlocality described above results in a situation where there are more than two wave solutions in the acoustic problem - such problems cannot be solved uniquely from the available acoustic boundary conditions. Nonlocality of the form:

\begin{align*}
i\omega v_i(x) = \int_\Omega\alpha_i(x-x')p(x')dx',\quad i\omega v_{i,i}(x)=\int_\Omega\beta(x-x') p(x')dx'
\end{align*}

essentially results in Eq. (\ref{eq:acousticWillis}) when Taylor series is employed and the first two terms are retained. We have already seen that the equi-frequency contours for (\ref{eq:acousticWillis}) are translated conics and that such an equation does correspond to a local acoustic media where interface problems are solvable through the usual interface conditions. If higher order terms in the Taylor series are non ignorable then the media cannot be approximated as a local media since the presence of higher order derivatives will, in general, give rise to more than two wave solutions which would render scattering problems unsolvable with the usual boundary conditions.

\subsection{Frequency regions of acceptable homogenized properties}

In addition to the presence of appropriate frequency dispersion in the form of equi-frequency surfaces, it is also important that the wave behavior is not Bragg dominated in the frequency regions where we wish to associate local properties to a composite. In considering the appropriate frequency region, we adopt the reasonable limits discussed for electromagnetism in Ref.\cite{smith2006homogenization} and for acoustics in Ref.\cite{ponge2017dynamic}. The main idea behind both these arguments is that the wavelength of the wave in the matrix of the composite should be roughly an order of magnitude larger than the representative cell length of the composite. If we denote the representative length by $a$ and the wavelength in the matrix by $\lambda$, then the first Bragg resonance in non-resonant composite occurs roughly at $a\approx \lambda/2$ \cite{simovski2007local}. In composites with resonant inclusions which might be candidates for coherent metamaterial properties, the resonant frequency will generally be lower than the Bragg resonance frequency. In general, we consider the region $a/\lambda\approx 0.001-0.01$ and the metamaterial frequency region $a/\lambda\approx 0.05-0.2$ \cite{simovski2009material}.

\section{Local acoustic metamaterial composites}

To see these effects in a composite, consider the 2D acoustic problem in a periodic composite. The acoustic wave equation is:

\begin{align*}
\nabla \cdot \left(\bar\rho\nabla p\right)=-\omega^2\bar\kappa p 
\end{align*}

where $\bar\rho=1/\rho,\bar\kappa=1/\kappa$. In index notation:

\begin{align*}
\bar\rho_{,i} p_{,i} + \bar\rho p_{,ii}=-\omega^2\bar\kappa p 
\end{align*}

Using Bloch theorem, we can expand $p=\bar{p}e^{i(k_1 x_1 + k_2 x_2)}=\bar{p}e^{ik_ix_i}$ where $\bar{p}$ is now unit cell periodic, we get:

\begin{align*}
\left[\bar\rho_{,i}(d/dx_i + ik_i) +  \bar\rho (d^2/dx_i^2+ 2ik_id/dx_i - k_i^2) \right]\bar p=-\omega^2\bar\kappa \bar p
\end{align*}

We can solve the above eigenvalue problem through a variety of techniques - here we use the Plane Wave method. We will not go into the details of this process as it is standard. We now discuss some equi-frequency contours which emerge for specific unit cells. The unit cell, $\Omega$, that we consider here is a three phase square unit cell with matrix region $\Omega_m$ and inclusion $\Omega_i$ (Fig. \ref{fig:w-solution}a). The third phase, $\Omega_t$, is a compliant phase which encapsulates the inclusion and allows it to resonate inside the unit cell. At some frequency $\omega$, the wavelength in the matrix is $\lambda_m=2\pi c_m/\omega$ where $c_m$ is the wave speed in the matrix material. If the unit cell length is $a$ then we will limit the metamaterial region to roughly $\lambda_m/a\approx 10$ which gives us $\omega\approx 2\pi c_m/10a$ - it is in this frequency region that the composite should exhibit local equi-frequency contours for it to appropriately be considered a local metamaterial. We define a normalized frequency $\bar{\omega}=a\omega/2\pi c_m$ and, thus, focus on the equi-frequency contours resulting in the range $\bar{\omega}\approx 0.1$. Fig. ({\color{blue}\ref{fig:w-solution}}a) shows the unit cell under consideration. The matrix material properties are taken to be those of water ($c_m=1500$m/s and $\rho_m=1000$kg/m$^3$), the inclusion is assumed to have the same properties whereas the shell medium has a bulk modulus which is hundred times smaller than water and a density that is half that of water. 

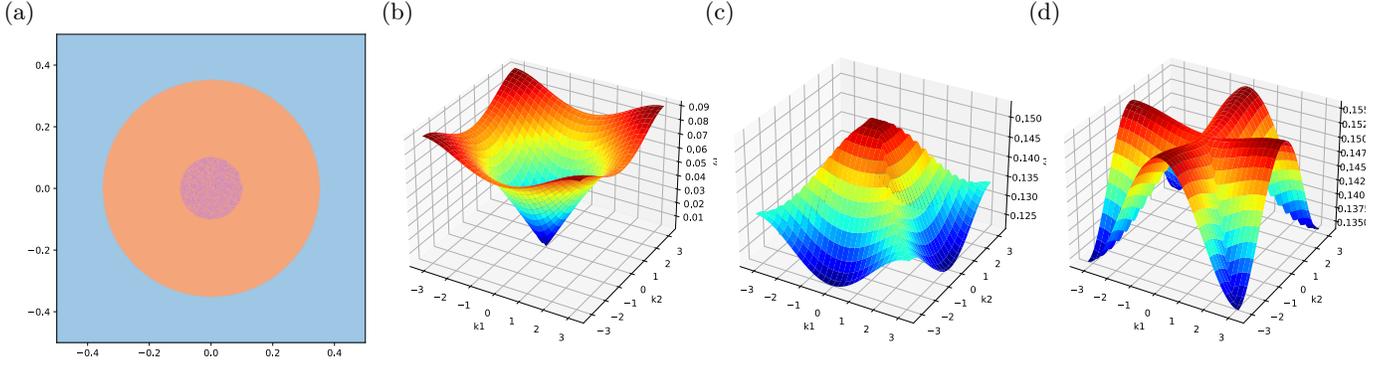
\begin{figure}[htp]
\centering
  \begin{overpic}[width=\textwidth,grid=false]{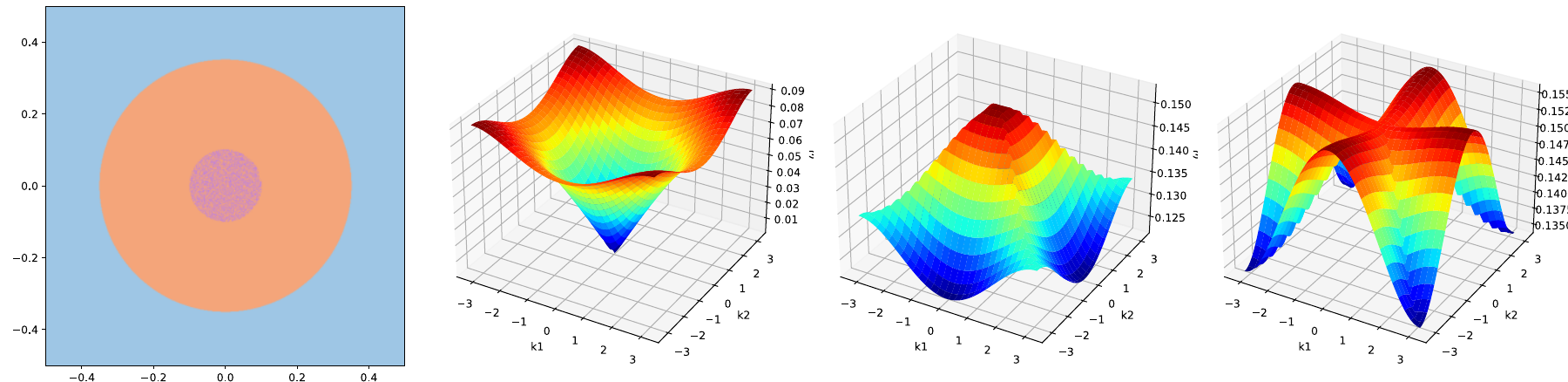}
    \put(-1,25){(a)}
    \put(27,25){(b)}
    \put(51,25){(c)}
    \put(75,25){(d)}
  \end{overpic}
\caption{Unit cell and eigen-branches, $\omega (k_1, k_2)$, of a 2D periodic composite. (a) Unit cell configuration. (b) First branch, (c) Second branch, (d) Third branch}
\label{fig:w-solution}
\end{figure}

Figs. (\ref{fig:w-solution}b-d) show the first three eigen-branches over the first Brillouin zone for the considered unit cell. In these sub-figures, the $z-$ axis is the normalized frequency $\bar{\omega}$. Due to the possibility of internal resonance, this is a unit cell which has the potential to exhibit negative effective properties on the second branch. For such properties to be local, as discussed above, there should be frequency regions where the EFCs can be approximated by conics. From Figs. (\ref{fig:w-solution}b,c), it's clear that there is a bandgap between the first and second branches and from Figs. (\ref{fig:w-solution}c,d), it is clear that there exists a frequency region on the second branch which also overlaps with the third branch. 
In this overlapping frequency region, there will be more that two wave solutions, in general, for certain wave propagation directions and, therefore, any approximation to conic will necessarily be partial.

\begin{figure}[htp]
\centering
  \begin{overpic}[width=\textwidth,grid=false]{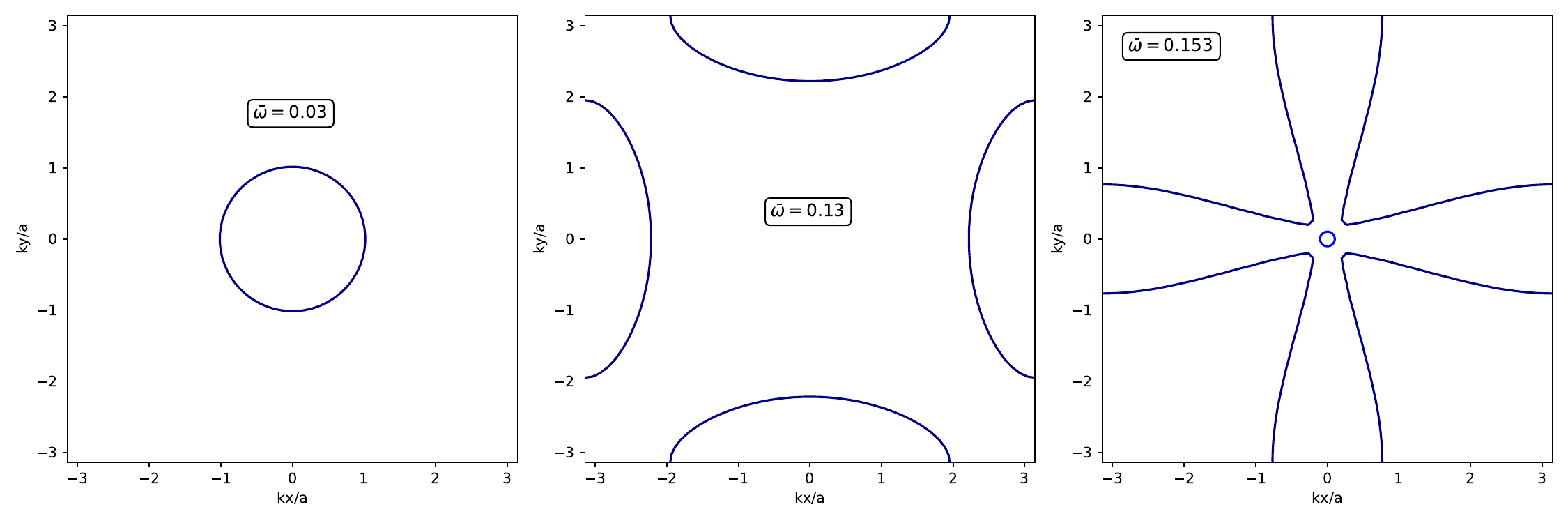}
    \put(0,33){(a)}
    \put(34,33){(b)}
    \put(67,33){(c)}
  \end{overpic}
\caption{Equi-frequency contours of the proposed 2D periodic composite for different normalized frequencies. (a) A circular EFC of the metamaterial at $\bar{\omega}=0.03$, where it behaves as a local metamaterial. (b) An EFC of the metamaterial at $\bar{\omega}=0.13$, where it can be considered local. (c) An EFC of the metamaterial at $\bar{\omega}=0.153$, where it exhibits partially local behavior.}
\label{fig:EFC-1}
\end{figure}

This is clarified in Fig. (\ref{fig:EFC-1}) where the equi-frequency contours are shown for three different normalized frequencies. The EFC at $\bar{\omega}=0.03$ is extracted from the first branch and is circular - clearly the composite behaves as an isotropic medium in this frequency region. Given the fact that $\nabla_\mathbf{k}\omega$ points outwards on the circular EFC, it is clear that in this frequency range the effective material properties are local, isotropic, and positive. The EFCs in the higher frequency ranges, however, point towards partially local anisotropic negative media.

Consider, for instance, the EFC at $\bar{\omega}=0.13$ which is shown in 
Fig. (\ref{fig:EFC-1}b). The EFC at this frequency resembles the EFCs in Figs. (\ref{fig:partially_local_metamaterial}). It is not difficult to extract a measure of the effective properties from the contour itself through a simple fitting process. The results are shown in Fig. (\ref{fig:contour_fit}) where two hyperbolas are used to fit parts of the contour in (Fig. \ref{fig:EFC-1}b). The effective material properties which emerge from this fitting process are also noted in these figures and they explicitly show that the entire contour in (Fig. \ref{fig:EFC-1}b) cannot be represented by the same material properties. Figs. (\ref{fig:contour_fit}a,b), together determine the range of directions in which the medium can be ascribed the two different sets of local material properties which are alluded to in the figure. As in Figs. (\ref{fig:partially_local_metamaterial}), no single conic can approximate the EFC at $\bar{\omega}=0.13$ and, therefore, the behavior of the composite is inherently non-local. It is possible to limit oneself to certain incidence directions and associated incident angle ranges such that the composite response is effectively local, in full correspondence to the discussion associated with Figs. (\ref{fig:partially_local_metamaterial}).  One relevant question here is: for what ranges of incident angles can we describe the response of the periodic structure using effective local properties? The answer to this question is found by considering the hyperbola in Fig. \ref{fig:contour_fit}a which intersects the $k_2a$ axis roughly at $k_2a=(-2,2)$. Therefore, as long as $-\sin^{-1}(2c_h/a\omega)\leq \theta\leq \sin^{-1}(2c_h/a\omega)$, the composite can be rigorously defined in terms of the local anisotropic effective properties which correspond to (Fig. \ref{fig:contour_fit}a). This range can be transformed into the incident angles-- $\theta\in[-\sin^{-1}(c_h/\pi c_m\bar{\omega}),\sin^{-1}(c_h/\pi c_m\bar{\omega})]$. If $c_h=c_m$ this range is larger than $-\pi/2,\pi/2$ at $\bar{\omega}=0.13$ which means that if the composite is interfaced with a homogeneous medium which has the same speed of sound as that of the matrix in the configuration shown in Fig. \ref{fig:second_third_overlap}a, then the composite can be rigorously defined by a single set of local effective material properties for the entire range of wave incidence angles at the considered interface.

Fig.~\ref{fig:EFC-1}c shows the equi-frequency contour at $\bar{\omega}=0.153$ which is another interesting case study in partially local behavior. At this frequency, there is an overlap between the second and third branches and this is evident in the complex EFC which appears in Fig.~\ref{fig:EFC-1}c. The central part of the EFC is nearly a circle -- without the contribution from the third branch, the composite could have been considered rigorously local at this frequency. However, the third branch contributes radiating EFC arms in Fig.~\ref{fig:EFC-1}c, which lead, in general, to more than 2 propagating solutions in a given direction. Over the circular part of the EFC, it can be shown that the group velocity, given by $\nabla_{k}\omega$, points inward on the circular equi-frequency contour (EFC), indicating negative refraction. This inward direction suggests that, if we could isolate the circular region from the rest of the EFC (for certain angles of incidence), then the effective material properties—such as bulk modulus or density— could reasonably take negative and isotropic values. A zoomed view of the circular part of the EFC is shown in Fig.~\ref{fig:second_third_overlap}c which confirms the circularity of the contour. In addition, it shows that if $k_2a\in[-0.1,0.1]$, then additional solutions corresponding to the third branch (in the form of the radiating arms of the RFC) do not show up. Therefore, if one were to have the interface as shown in Fig.~(\ref{fig:second_third_overlap}a) and if one were to limit the incidence angles so that $k_2a\in[-0.1,0.1]$, then at $\bar{\omega}=0.153$ the composite behaves like a isotropic local metamaterial. Outside this window, the hyperbolic branch intrudes and the local description breaks down, so one cannot use a single set of effective parameters over the full incidence range.  Consequently, the admissible incidence angles are confined to
\begin{equation}
\theta \in \bigl[-\sin^{-1}\!\bigl(\tfrac{1}{20\pi\bar\omega}\bigr)\,,\,
               \sin^{-1}\!\bigl(\tfrac{1}{20\pi\bar\omega}\bigr)\bigr]
\;\approx\;[-5^\circ,5^\circ]
\quad(\bar\omega=0.153),
\end{equation}
beyond which the scattering coefficients must be computed using a nonlocal description. 

\begin{figure}[htp]
\centering
  \begin{overpic}[width=\textwidth,grid=false]{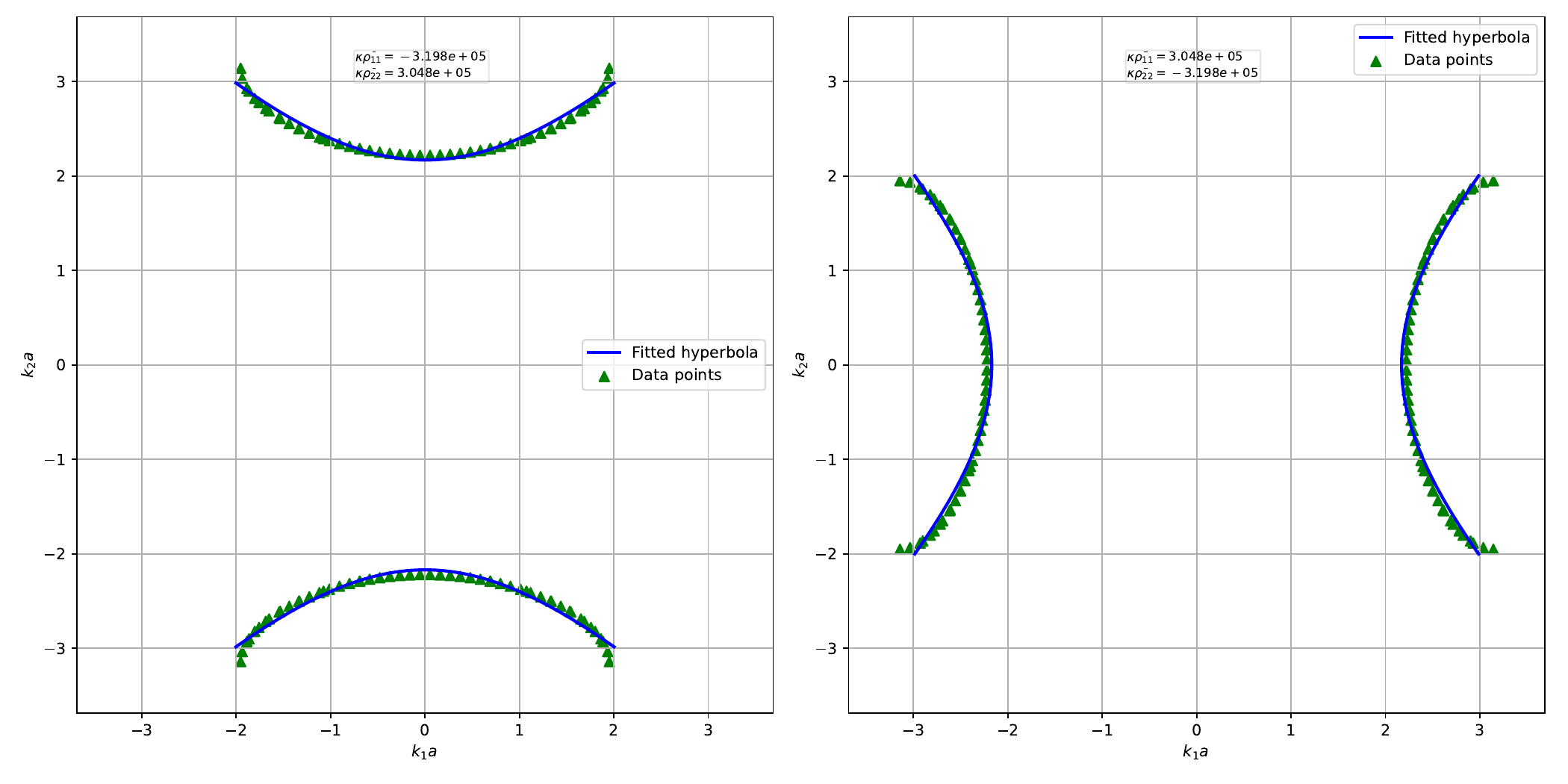}
    \put(1,50){(a)}
    \put(51,50){(b)}
  \end{overpic}
\caption{ (a) A partial contour fitted with a hyperbola, showing the effective material properties where $\kappa\bar{\rho_{11}}>0$ and $\kappa\bar{\rho_{22}}<0$. (b) Another partial contour fitted with a different hyperbola, illustrating the effective material properties where $\kappa\bar{\rho_{22}}>0$ and $\kappa\bar{\rho_{11}}<0$. }
\label{fig:contour_fit}
\end{figure}

 \begin{figure}[h]
\centering
\begin{overpic}[width=\textwidth,grid=false]{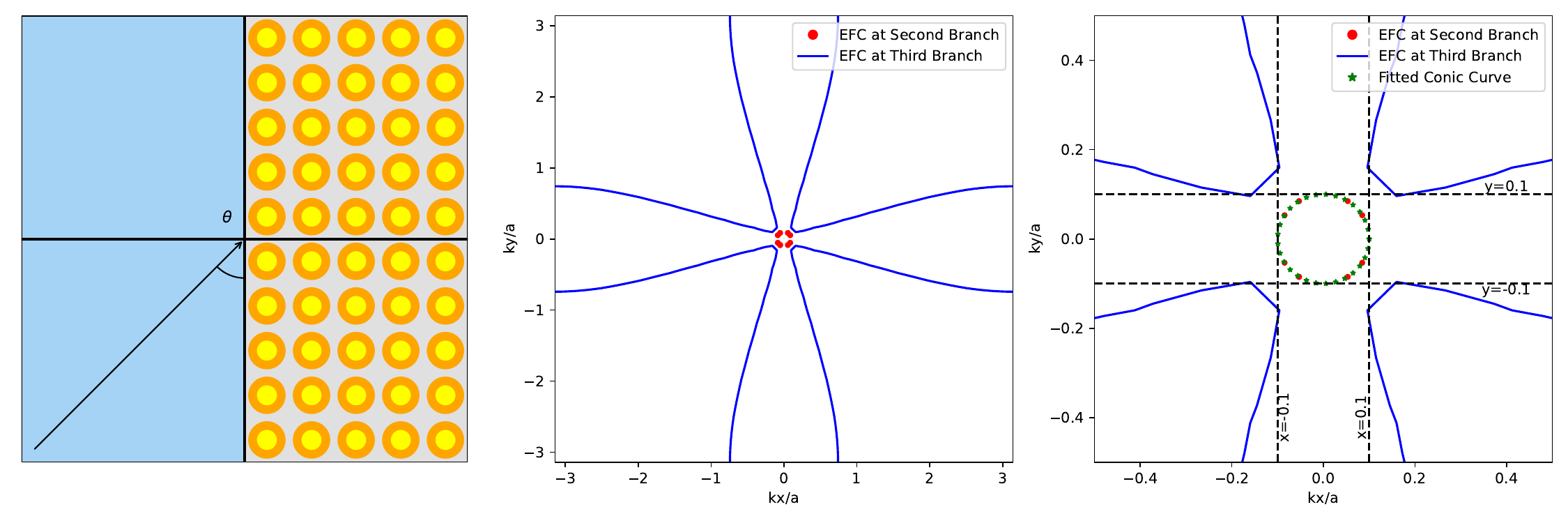}
    \put(-2,33){(a)}
    \put(31,33){(b)}
    \put(66,33){(c)}
    
  \end{overpic}
  \caption{ (a) Scattering problem at an interface between a homogeneous medium and a 2D composite.(a) EFC at $\bar\omega=0.153$, red points show the second-branch EFC and the blue line is located in the third-branch; (b) zoomed view EFC at $\bar\omega=0.153$ showing green stars for the fitted conic curve where the metamaterial behaves locally.}
\label{fig:second_third_overlap}
\end{figure}

\subsection{Effective properties through field averaging}\label{some_com_eff}

We can calculate the effective properties of the composite by using various techniques elaborated upon in the literature. For the purpose of discussion, here we use the field averaging technique~\cite{willis2009exact, amirkhizi2017homogenization, nemat2011overall, srivastava2012overall}, which automatically satisfies the dispersion relation (as other carefully designed techniques also do). We focus on the cases shown in Fig.\ref{fig:EFC-1} with $\bar{\omega}=0.03,0.153$. We denote the effective properties by effective bulk modulus $\tilde{\kappa}$ and effective density components $\tilde{\rho}_{ij}$. 
Because the equi-frequency contour in Fig.~\ref{fig:EFC-1}a is perfectly circular, the off-diagonal inertial coupling vanishes, i.e.\ $\tilde\rho_{12}=\tilde\rho_{21}=0$.  Furthermore, at the second-branch frequency $\bar\omega=0.153$ the EFC remains circular within a small wavenumber region, indicating locally acoustic behavior there as well.   
Recalling the acoustic constitutive relations but writing them now in their averaged form:

\begin{align}\label{eq:homogenized}
i\omega V_1 = \frac{1}{\tilde{\rho}_{11}}P_{,1},\quad i\omega V_2 = \frac{1}{\tilde{\rho}_{22}}P_{,2},\quad i\omega P=\tilde{\kappa} V_{i,i}
\end{align}

where the field variables are unit cell averaged variables of the form $S=\bar{S}\exp(ik_ix_i)$. The constant term $\bar{S}$ is computed by averaging the unit cell periodic part of the corresponding microscale variable $s(x_1,x_2)$ which, as per earlier convention, is given by $\bar{s}(x_1,x_2)$. Representing unit cell average through $\langle.\rangle_\Omega$, we have $\bar{V}_m=\langle v_m\exp(-ik_ix_i)\rangle_\Omega$ and $\bar{P}_{,m}=\langle p_{,m}\exp(-ik_ix_i)\rangle_\Omega$, and thus:

\begin{align*}
\tilde{\rho}_{11} = \frac{\langle\bar{p}_{,1}+ik_1\bar{p}\rangle_\Omega}{i\omega\langle\bar{v}_1\rangle_\Omega},\quad \tilde{\rho}_{22} = \frac{\langle\bar{p}_{,2}+ik_2\bar{p}\rangle_\Omega}{i\omega\langle\bar{v}_2\rangle_\Omega},\quad \tilde{\kappa} = \frac{i\omega\langle\bar{p}\rangle_\Omega}{\langle\bar{v}_{1,1}+ik_1\bar{v}_1\rangle_\Omega + \langle\bar{v}_{2,2}+ik_2\bar{v}_2\rangle_\Omega}
\end{align*}

When we calculate these properties at $\bar{\omega}=0.03$ and for different points on the EFC shown in Fig. \ref{fig:EFC-1}a, we find that we get very similar effective properties everywhere. These values are $\tilde{\rho}_{11}=\tilde{\rho}_{22}\approx 713$ kg/m$^3$ and $\tilde{\kappa}\approx 5.54\times10^7$ Pa. These properties satisfy the following effective dispersion relation to order $\mathcal{O}(10^{-3})$:

\begin{align*}\label{eq:EFCacoustics2D2}
\frac{\tilde{\kappa}}{\omega^2\tilde{\rho}_{11}}k_1^2 + \frac{\tilde{\kappa}}{\omega^2\tilde{\rho}_{22}}k_2^2 -1=0
\end{align*}
Thus we are able to determine a single set of isotropic effective properties through field averaging which represents the effectively isotropic wave response of the composite at $\bar{\omega}=0.03$.

In contrast to the discussion in the low frequency regime where the EFC is a circle, when the effective properties are computed through field averaging for various points in Fig. \ref{fig:EFC-1}b (corresponding to $\bar{\omega}=0.13$), we find that these properties vary widely. In this regime, the effective properties are capturing the non-locality which is inherent in the double hyperbolas of the EFC. Since we would like to represent the two hyperbolas separately, as discussed above, effective property extraction through usual techniques such as averaging does not yield the desired results. The computed effective properties at \(\bar{\omega}=0.153\) are \(\tilde{\rho}\approx -3.3418\ \mathrm{kg/m^3}\) and  \(\tilde{\kappa}\approx -4.7048\times10^8\ \mathrm{Pa}\) indicating negative effective properties.

\section{Transition layers}

We now want to use these effective properties (defined for a restricted region where the composite behaves locally) to solve a scattering problem for an appropriate range of angles. The idea is shown in Fig. \ref{fig:scattering-transition}. Fig. \ref{fig:scattering-transition}a shows a scattering problem at an interface between a known homogeneous medium with material properties $\kappa_h,\rho_h$ and a composite medium. The scattering problem is solved in a frequency regime where we expect the composite to behave locally. For acoustic problems, this has the effect of ensuring that there is only a single transmitted propagating wave and a single reflected propagating wave. This does not mean that there are no other waves at the interface. In fact, there is an interface region which is populated by a host of non-propagating and evanescent wave modes and these modes are necessary to satisfy the complicated interface continuity conditions which exist at the interface. Once the scattering problem is solved for the homogeneous-composite interface, the interface is replaced with a transition layer (shown in gray in Fig. \ref{fig:scattering-transition}b) and the bulk of the composite is replaced with a homogeneous medium. The transition layer is necessary in order to approximate the region near the interface which consists of evanescent wave modes - if this region is to be represented through some overall properties then these properties certainly cannot be the same as those which are in the bulk of the composite. The appropriate thickness of this layer -- $t$ -- has been the subject of long debate in the electromagnetism community but a good choice of this thickness is $a$ \cite{simovski2009material,drude1925theory}, which is the length scale of the unit cell which forms the composite. The properties of the transition layer are denoted by $\kappa_t,\boldsymbol{\rho}_t$ whereas those of the equivalent homogeneous medium are denoted by $\tilde{\kappa},\tilde{\boldsymbol{\rho}}$, in congruence with earlier defined notation.

We aim to show that in the region of locality, the scattering problem can be solved using just the effective properties in a range of angles and the results agree well with the scattering calculations of the original problem.

\begin{figure}[htp]
\centering
\begin{overpic}[width=\textwidth,grid=false]{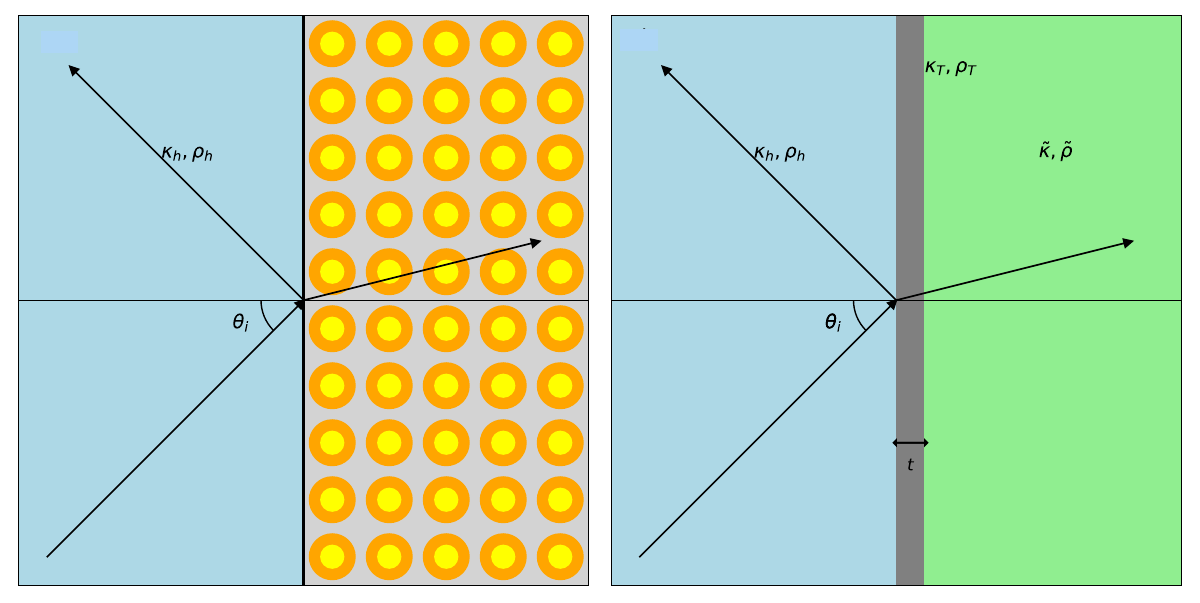}
    \put(5,3){(a)}
    \put(55,3){(b)}
  \end{overpic}
\caption{(a) Scattering problem at an interface between a homogeneous medium and a 2D composite material with angles of incidents $\theta$. (b) Scattering problem between a homogeneous material and the homogenized composite material where the interface replaced with a transition layer.}
\label{fig:scattering-transition}
\end{figure}

\subsection{Effective properties based upon averaging}

One route to determining the local effective properties is to assign the unit cell averaged properties (Eq.~\ref{eq:homogenized}) to $\tilde{\kappa},\tilde{\boldsymbol{\rho}}$ as they satisfy the dispersion relation and are consistently defined through a dynamic homogenization procedure \cite{willis2009exact, amirkhizi2017homogenization, nemat2011overall, srivastava2012overall}. Once this is done, the effective properties of the transition layer may be determined as an average of material properties of the homogeneous medium and the effective one \cite{drude1925theory}, and its thickness may be taken as the unit cell length scale $a$ \cite{simovski2009material, drude1925theory}. We adopt this thickness because Drude’s local-field analysis shows that dipoles one unit cell beneath the surface already exhibit bulk-like Clausius–Mossotti behavior, whereas surface dipoles do not\cite{simovski2009material, drude1925theory}. Although the exact thickness is not universally defined~\cite{simovski2000dielectric,simovski2002bulk}, treating the Drude layer as a unit-cell‐sized fitting parameter is particularly effective for metamaterial slabs comprising only a few cells. Importantly, in the quasi-static or long-wavelength regime (\(\lambda \gg a\)), the Drude layer cannot enforce correct boundary behavior because the accumulated phase per cell is non-negligible; nonetheless, its inclusion yields more accurate reflection and transmission predictions while preserving the simplicity of local, wavevector-independent effective parameters. This approach should work reasonably well in those frequency ranges where the EFC is circular. For cases where the EFC is doubly-hyperbolic (such as Fig.~\ref{fig:EFC-1}b), this process will not work since dynamic homogenization will not automatically result in local effective properties (since only singly-hyperbolic EFCs are signature of local properties). Here, we check the applicability of the effective properties in the bulk and the averaged properties for the transition layer by comparing scattering calculations for the two situations shown in Fig.~\ref{fig:scattering-transition}. This process involves solving the scattering problem for two cases: one involving two interfaces between three homogeneous mediums, and second involving a single interface between a homogeneous medium and a phononic crystal. At $\bar{\omega}=0.153$, the metamaterial’s response is isotropic, so dynamic homogenization yields an isotropic density tensor $\tilde{\rho}$, and the effective parameters reduce to the scalars $\rho_t$, $\kappa_t$, and $c_t$.

We perform \texttt{k-Wave} pseudo-spectral time-domain simulations \cite{treeby2010k} over the range of admissible incidence angles \(\theta\) within which the composite can be described through local properties. In our simulation, the scattering setup uses a planar (line) pressure source driven by a one-cycle tone burst—i.e.\ a finite sinusoidal pulse zero-padded to the full simulation duration.  Pressure sensors are placed both in the homogeneous host medium and within the metamaterial; by Fourier-transforming these time-domain recordings, we extract high-fidelity “ground-truth” reflection and transmission coefficients \(R(\theta)\) and \(T(\theta)\).

At $\bar{\omega}=0.03$, the simulation results are trivial since the EFC is circular and the composite behaves locally over all angles of incidences - this is also the region where the evanescent modes are not prominent (since the wavelength is much larger than the heterogeneity length scale). This means that transition layers do not appreciably improve the already good correspondence between the scattering results from the original problem and the homogenized problem.

At $\bar{\omega}=0.153$, we compare two simulations: 

\begin{enumerate}
    \item A homogeneous effective‐medium model with parameters \((\tilde{\kappa},\tilde{\rho})\) but no transition layer.
    \item The same homogenized medium augmented by a Drude transition layer of thickness \(\Delta = a\), whose parameters \((\kappa_t,\rho_t)\) are averages of the properties of the homogeneous medium to the left and the homogenized medium to the right.
\end{enumerate}

By comparing these two cases, we demonstrate the practical benefit of a simple, one‐cell‐thick layer: it captures all significant evanescent contributions without requiring costly integral‐equation or multi‐mode expansions. These simulations are done for a range of incidence angles over which we have already determined the composite to act locally.

\begin{figure}[h]
   \centering
\begin{overpic}[width=\textwidth,grid=false]{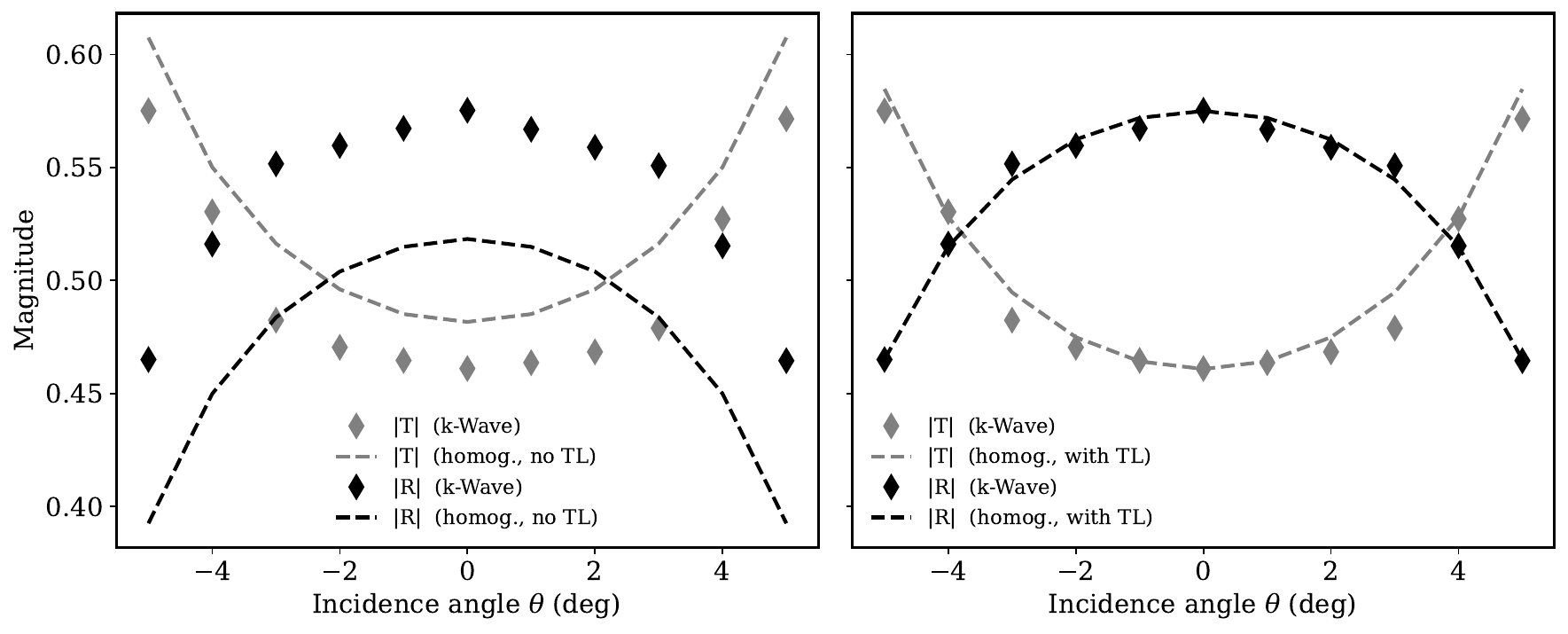}
    \put(0,40){(a)}
    \put(52,40){(b)}
  \end{overpic}
  \caption{(a) Scattering at $\bar\omega = 0.153$ using a homogeneous model (no transition layers). Reflection $|R|$ (black) and transmission $|T|$ (gray) vs. incidence angle. Diamonds: k-Wave results. Deviation at different angles shows missing evanescent effects. (b) Same setup with Drude layers of thickness $\Delta = a$. Dashed lines: Drude model results.}
  \label{fig:scattering}
\end{figure}

Figure~\ref{fig:scattering}a shows a comparison between the scattering parameters computed for the original scattering problem and the homogenized problem with no transition layer. It shows that the transmission and reflection coefficients predicted by the homogenized model deviate from the scattering pramaters of the original problem across the range of incidence angles. This discrepancy arises because evanescent boundary modes—which are strong at the normalized frequency $\bar\omega=0.153$ are not captured by a sharp interface between the homogeneous medium and the homogenized medium. The resulting errors exceed 20\% in reflection for angles beyond 4$^\circ$. Figure~\ref{fig:scattering}b shows that introducing a one‐cell‐thick Drude transition layer restores agreement between the results of the original scattering problem and the homogenized scattering problem. The errors in this case are below 2\% across the range of incidence angles considered. This confirms that a thin artificial transition layer of one unit-cell thickness suffices to represent evanescent‐mode contributions in local acoustic metamaterials.

\section{Conclusions}

In this paper, we begin by characterizing local acoustic metamaterials via their equi‐frequency contours (EFCs). For multiple plane waves interacting at an interface, the three continuity conditions (pressure and normal velocity) are sufficient to determine all scattering parameters uniquely only when exactly two plane waves—in addition to the incident wave—are involved, one in each medium (the reflected and transmitted waves). Decomposing the wavevector into a component parallel to the interface, \(\mathbf{k}_{\parallel}\), and a perpendicular component, \(k_{\perp}\), this requirement is equivalent to demanding that for a given \(\omega\) and \(\mathbf{k}_{\parallel}\), there exists a single real \(k_{\perp}\) in each medium whose energy flux is either away from the interface or fully along it; equivalently, the EFC must trace a conic section. To realize this local‐medium behavior in both the first branch and a narrow band of the second branch, we introduce a novel square unit cell comprising a resonant inclusion surrounded by a shell. Applying the field‐averaging homogenization method to this design yields negative effective parameters \(\tilde{\kappa}\) and \(\tilde{\rho}\) at the normalized frequency \(\bar{\omega}=0.153\).

To validate scattering predictions, we simulate an interface between a homogeneous background and our composite using the k‐Wave MATLAB toolbox. We record time‐domain pressures on sensors in each region, Fourier‐transform at \(\bar{\omega}=0.153\), and extract the complex amplitudes \(P_{\rm inc}\), \(P_{\rm rx}\), and \(P_{\rm tx}\) to compute the reflection and transmission coefficients \(R(\theta)\) and \(T(\theta)\). While sharp‐interface homogenization alone incurs reflection errors above \(20\%\) at oblique angles—due to uncaptured evanescent‐mode contributions—a one‐cell‐thick Drude transition layer fitted at normal incidence restores agreement with full‐wave simulations to within \(2\%\) for all \(\theta\). Crucially, once the Drude layer’s density and bulk modulus are determined, any further reflection or transmission calculation reduces to enforcing continuity at two boundaries—no additional full‐wave solves are needed—making our approach highly efficient for rapid parameter sweeps, optimization loops, and integration into the design of acoustic lenses, cloaks, and waveguides.

\section{Acknowledgments}\label{sec:ack}
A.S. acknowledges support from the NSF grant \#2219203 to the Illinois Institute of Technology.



\begin{thebibliography}{10}

\bibitem{placidi2016review}
Luca Placidi, Emilio Barchiesi, Emilio Turco, and Nicola~Luigi Rizzi.
\newblock A review on 2d models for the description of pantographic fabrics.
\newblock {\em Zeitschrift f{\"u}r angewandte Mathematik und Physik}, 67:1--20,
  2016.

\bibitem{barchiesi2019pantographic}
Emilio Barchiesi, Simon~R Eugster, Luca Placidi, and Francesco Dell’Isola.
\newblock Pantographic beam: a complete second gradient 1d-continuum in plane.
\newblock {\em Zeitschrift f{\"u}r angewandte Mathematik und Physik}, 70:1--24,
  2019.

\bibitem{Nemat-Nasser2011HomogenizationMaterials}
Sia Nemat-Nasser, John~R Willis, Ankit Srivastava, and Alireza~V Amirkhizi.
\newblock {Homogenization of periodic elastic composites and locally resonant
  sonic materials}.
\newblock {\em Physical Review B - Condensed Matter and Materials Physics},
  83(10), 2011.

\bibitem{Nemat-Nasser2011OverallComposites}
S.~Nemat-Nasser and A.~Srivastava.
\newblock {Overall dynamic constitutive relations of layered elastic
  composites}.
\newblock {\em Journal of the Mechanics and Physics of Solids}, 59(10), 2011.

\bibitem{Srivastava2011OverallComposites}
A.~Srivastava and S.~Nemat-Nasser.
\newblock {Overall dynamic constitutive relations for layered elastic
  composites}.
\newblock In {\em Proceedings of SPIE - The International Society for Optical
  Engineering}, volume 7978, 2011.

\bibitem{bensoussan2011asymptotic}
Alain Bensoussan, Jacques-Louis Lions, and George Papanicolaou.
\newblock {\em Asymptotic analysis for periodic structures}, volume 374.
\newblock American Mathematical Soc., 2011.

\bibitem{sanchez1980non}
Enrique S{\'a}nchez-Palencia.
\newblock Non-homogeneous media and vibration theory.
\newblock {\em Lecture Note in Physics, Springer-Verlag}, 320:57--65, 1980.

\bibitem{bakhvalov2012homogenisation}
Nikolai~Sergeevich Bakhvalov and Grigory Panasenko.
\newblock {\em Homogenisation: averaging processes in periodic media:
  mathematical problems in the mechanics of composite materials}, volume~36.
\newblock Springer Science \& Business Media, 2012.

\bibitem{parnell2006dynamic}
William~J Parnell and I~David Abrahams.
\newblock {Dynamic homogenization in periodic fibre reinforced media.
  Quasi-static limit for SH waves}.
\newblock {\em Wave Motion}, 43(6):474--498, 2006.

\bibitem{andrianov2008higher}
Igor~V Andrianov, Vladimir~I Bolshakov, Vladyslav~V Danishevs'~kyy, and Dieter
  Weichert.
\newblock Higher order asymptotic homogenization and wave propagation in
  periodic composite materials.
\newblock {\em Proceedings of the Royal Society A: Mathematical, Physical and
  Engineering Sciences}, 464(2093):1181--1201, 2008.

\bibitem{craster2009mechanism}
RV~Craster, S{\'e}bastien Guenneau, and SDM Adams.
\newblock Mechanism for slow waves near cutoff frequencies in periodic
  waveguides.
\newblock {\em Physical Review B}, 79(4):045129, 2009.

\bibitem{craster2010high}
R~V Craster, J~Kaplunov, and A~V Pichugin.
\newblock {High-frequency homogenization for periodic media}.
\newblock {\em Proceedings of the Royal Society A: Mathematical, Physical and
  Engineering Science}, 466(2120):2341--2362, 2010.

\bibitem{antonakakis2013asymptotics}
T~Antonakakis, Richard~V Craster, and S{\'e}bastien Guenneau.
\newblock Asymptotics for metamaterials and photonic crystals.
\newblock {\em Proceedings of the Royal Society A: Mathematical, Physical and
  Engineering Sciences}, 469(2152):20120533, 2013.

\bibitem{antonakakis2014homogenisation}
T~Antonakakis, Richard~V Craster, and S{\'e}bastien Guenneau.
\newblock Homogenisation for elastic photonic crystals and dynamic anisotropy.
\newblock {\em Journal of the Mechanics and Physics of Solids}, 71:84--96,
  2014.

\bibitem{willis2009exact}
John~R Willis.
\newblock {Exact effective relations for dynamics of a laminated body}.
\newblock {\em Mechanics of Materials}, 41(4):385--393, 2009.

\bibitem{amirkhizi2017homogenization}
Alireza~V. Amirkhizi.
\newblock {Homogenization of layered media based on scattering response and
  field integration}.
\newblock {\em Mechanics of Materials}, 114:76--87, 11 2017.

\bibitem{nemat2011overall}
Sia Nemat-Nasser and Ankit Srivastava.
\newblock {Overall dynamic constitutive relations of layered elastic
  composites}.
\newblock {\em Journal of the Mechanics and Physics of Solids},
  59(10):1953--1965, 2011.

\bibitem{srivastava2012overall}
Ankit Srivastava and Sia Nemat-Nasser.
\newblock {Overall dynamic properties of three-dimensional periodic elastic
  composites}.
\newblock {\em Proceedings of the Royal Society A: Mathematical, Physical and
  Engineering Science}, 468(2137):269--287, 2012.

\bibitem{willis1981variational}
John~R Willis.
\newblock Variational and related methods for the overall properties of
  composites.
\newblock {\em Advances in applied mechanics}, 21:1--78, 1981.

\bibitem{willis1983overall}
John~R Willis.
\newblock {The overall elastic response of composite materials}.
\newblock {\em Journal of Applied Mechanics}, 50:1202--1209, 1983.

\bibitem{willis1984variational}
JR~Willis.
\newblock Variational principles and operator equations for electromagnetic
  waves in inhomogeneous media.
\newblock {\em Wave Motion}, 6(2):127--139, 1984.

\bibitem{shuvalov2011effective}
AL~Shuvalov, AA~Kutsenko, AN~Norris, and O~Poncelet.
\newblock Effective willis constitutive equations for periodically stratified
  anisotropic elastic media.
\newblock {\em Proceedings of the Royal Society A: Mathematical, Physical and
  Engineering Sciences}, 467(2130):1749--1769, 2011.

\bibitem{willis2011effective}
John~R Willis.
\newblock Effective constitutive relations for waves in composites and
  metamaterials.
\newblock {\em Proceedings of the Royal Society A: Mathematical, Physical and
  Engineering Sciences}, 467(2131):1865--1879, 2011.

\bibitem{norris2012analytical}
A~N Norris, A~L Shuvalov, and A~A Kutsenko.
\newblock {Analytical formulation of three-dimensional dynamic homogenization
  for periodic elastic systems}.
\newblock {\em Proceedings of the Royal Society A: Mathematical, Physical and
  Engineering Science}, 468(2142):1629--1651, 2012.

\bibitem{milton2007modifications}
Graeme~W Milton and John~R Willis.
\newblock {On modifications of Newton's second law and linear continuum
  elastodynamics}.
\newblock {\em Proceedings of the Royal Society A: Mathematical, Physical and
  Engineering Science}, 463(2079):855, 2007.

\bibitem{srivastava2015elastic}
Ankit Srivastava.
\newblock Elastic metamaterials and dynamic homogenization: a review.
\newblock {\em International Journal of Smart and Nano Materials}, 6(1):41--60,
  2015.

\bibitem{simovski2009material}
C~R Simovski.
\newblock {Material parameters of metamaterials (a review)}.
\newblock {\em Optics and Spectroscopy}, 107(5):726--753, 2009.

\bibitem{el2000metallic}
I~El-Kady, MM~Sigalas, R~Biswas, KM~Ho, and CM~Soukoulis.
\newblock Metallic photonic crystals at optical wavelengths.
\newblock {\em Physical Review B}, 62(23):15299, 2000.

\bibitem{smith2002determination}
D~Robert Smith, Sheldon Schultz, P~Marko{\v{s}}, and Costas~M Soukoulis.
\newblock Determination of effective permittivity and permeability of
  metamaterials from reflection and transmission coefficients.
\newblock {\em Physical review B}, 65(19):195104, 2002.

\bibitem{amirkhizi2018overall}
Alireza~V Amirkhizi and Vahidreza Alizadeh.
\newblock Overall constitutive description of symmetric layered media based on
  scattering of oblique sh waves.
\newblock {\em Wave Motion}, 83:214--226, 2018.

\bibitem{kunin2012elastic}
Isaak~Abramovich Kunin.
\newblock {\em Elastic media with microstructure I: one-dimensional models},
  volume~26.
\newblock Springer Science \& Business Media, 2012.

\bibitem{cordaro2023solving}
Andrea Cordaro, Brian Edwards, Vahid Nikkhah, Andrea Al{\`u}, Nader Engheta,
  and Albert Polman.
\newblock Solving integral equations in free space with inverse-designed
  ultrathin optical metagratings.
\newblock {\em Nature Nanotechnology}, 18(4):365--372, 2023.

\bibitem{hatamzadeh2011numerical}
Saeed Hatamzadeh-Varmazyar and Zahra Masouri.
\newblock Numerical method for analysis of one-and two-dimensional
  electromagnetic scattering based on using linear fredholm integral equation
  models.
\newblock {\em Mathematical and Computer Modelling}, 54(9-10):2199--2210, 2011.

\bibitem{eringen2012microcontinuum}
A~Cemal Eringen.
\newblock {\em Microcontinuum field theories: I. Foundations and solids}.
\newblock Springer Science \& Business Media, 2012.

\bibitem{eringen2001microcontinuum}
A~Cemal Eringen.
\newblock {\em Microcontinuum field theories: II. Fluent media}, volume~2.
\newblock Springer Science \& Business Media, 2001.

\bibitem{eringen1983differential}
A~Cemal Eringen.
\newblock On differential equations of nonlocal elasticity and solutions of
  screw dislocation and surface waves.
\newblock {\em Journal of applied physics}, 54(9):4703--4710, 1983.

\bibitem{wang2010micromorphic}
Xianqiao Wang and James~D Lee.
\newblock Micromorphic theory: a gateway to nano world.
\newblock {\em International Journal of Smart and Nano Materials},
  1(2):115--135, 2010.

\bibitem{norris2009acoustic}
Andrew~N Norris.
\newblock {Acoustic metafluids}.
\newblock {\em The Journal of the Acoustical Society of America},
  125(2):839--849, 2009.

\bibitem{norris2011elastic}
A~N Norris and A~L Shuvalov.
\newblock {Elastic cloaking theory}.
\newblock {\em Wave Motion}, 48(6):525--538, 2011.

\bibitem{srivastava2014limit}
Ankit Srivastava and Sia Nemat-Nasser.
\newblock On the limit and applicability of dynamic homogenization.
\newblock {\em Wave Motion}, 51(7):1045--1054, 2014.

\bibitem{willis2013some}
John~R Willis.
\newblock {Some thoughts on dynamic effective properties--a working document}.
\newblock {\em arXiv preprint arXiv:1311.3875}, 2013.

\bibitem{joseph2015reflection}
L~M Joseph and R~V Craster.
\newblock {Reflection from a semi-infinite stack of layers using
  homogenization}.
\newblock {\em Wave Motion}, 54:145--156, 2015.

\bibitem{drude1925theory}
Paul Drude.
\newblock {\em {The theory of optics}}.
\newblock Courier Corporation, 1925.

\bibitem{strachan1933reflexion}
C~Strachan.
\newblock {The reflexion of light at a surface covered by a monomolecular
  film}.
\newblock In {\em Mathematical Proceedings of the Cambridge Philosophical
  Society}, volume~29, pages 116--130. Cambridge Univ Press, 1933.

\bibitem{srivastava2017evanescent}
Ankit Srivastava and John~R Willis.
\newblock Evanescent wave boundary layers in metamaterials and sidestepping
  them through a variational approach.
\newblock {\em Proceedings of the Royal Society A: Mathematical, Physical and
  Engineering Sciences}, 473(2200):20160765, 2017.

\bibitem{treeby2010k}
Bradley~E Treeby and Benjamin~T Cox.
\newblock k-wave: Matlab toolbox for the simulation and reconstruction of
  photoacoustic wave fields.
\newblock {\em Journal of biomedical optics}, 15(2):021314--021314, 2010.

\bibitem{smith2006homogenization}
David~R Smith and John~B Pendry.
\newblock Homogenization of metamaterials by field averaging.
\newblock {\em JOSA B}, 23(3):391--403, 2006.

\bibitem{ponge2017dynamic}
Marie-Fraise Ponge, Olivier Poncelet, and Daniel Torrent.
\newblock Dynamic homogenization theory for nonlocal acoustic metamaterials.
\newblock {\em Extreme Mechanics Letters}, 12:71--76, 2017.

\bibitem{simovski2007local}
Constantin~R Simovski and Sergei~A Tretyakov.
\newblock Local constitutive parameters of metamaterials from an
  effective-medium perspective.
\newblock {\em Physical Review B—Condensed Matter and Materials Physics},
  75(19):195111, 2007.

\bibitem{simovski2000dielectric}
Constantin~R Simovski, Mikhail Popov, and Sailing He.
\newblock Dielectric properties of a thin film consisting of a few layers of
  molecules or particles.
\newblock {\em Physical Review B}, 62(20):13718, 2000.

\bibitem{simovski2002bulk}
CR~Simovski and B~Sauviac.
\newblock On the bulk averaging approach for obtaining the effective parameters
  of thin magnetic granular films.
\newblock {\em The European Physical Journal-Applied Physics}, 17(1):11--20,
  2002.

\end{thebibliography}

\end{document}